\documentclass[aps,superscriptaddress,floatfix]{revtex4}
\usepackage{graphicx}
\usepackage{epsfig}
\usepackage{amsmath,amssymb}
\usepackage{color}
\begin{document}
\newcommand{\be}{\begin{eqnarray}}
\newcommand{\ee}{\end{eqnarray}}
\newcommand\del{\partial}
\newcommand\nn{\nonumber}
\newcommand{\Tr}{{\rm Tr}}
\newcommand{\Trg}{{\rm Str}}
\newcommand{\mat}{\left ( \begin{array}{cc}}
\newcommand{\emat}{\end{array} \right )}
\newcommand{\matt}{\left ( \begin{array}{ccc}}
\newcommand{\ematt}{\end{array} \right )}
\newcommand{\vect}{\left ( \begin{array}{c}}
\newcommand{\evect}{\end{array} \right )}
 \definecolor{Bittersweet}   {cmyk}{0,0.75,1,0.24}
\newcommand{\rbs}{\color{Bittersweet}}

\voffset 0cm
\hoffset 0cm

\title{Nonhermitian Supersymmetric Partition Functions: the case
of one bosonic flavor}

\author{K.\ Splittorff}
\affiliation{The Niels Bohr Institute, Blegdamsvej 17, DK-2100,
Copenhagen {\O}, Denmark}
\author{J.J.M.\ Verbaarschot$^{1,\, }$}
\affiliation{Department of Physics and Astronomy, SUNY, Stony Brook,
New York 11794, USA}
\author{M.R.\ Zirnbauer}
\affiliation{Institut f\"ur Theoretische Physik, Universit\"at zu
K\"oln, Z\"ulpicher Stra{\ss}e 77, 50937 K\"oln, Germany.}
\date{Feb 15, 2008}
\begin{abstract} We discuss the supersymmetric formulation of the
nonhermitian $\beta = 2$ random matrix partition function with one
bosonic flavor. This partition function is regularized by adding one
conjugate boson and fermion each. A supersymmetric nonlinear
$\sigma$-model for the resulting Goldstone degrees of freedom is
obtained using symmetry arguments only. For a Gaussian probability
distribution the same results are derived using superbosonization and
the complex orthogonal polynomial method. The symmetry arguments
apply to any model with the same symmetries and a mass gap, and
demonstrate the universality of the nonlinear $\sigma$-model.
\end{abstract}
\maketitle

\section{Introduction}

There exists a vast literature \cite{GM-GW} showing that the spectra
of many physical systems on the scale of the average level spacing
(the \emph{microscopic} scale) are correlated according to universal
laws given by random matrix theory. They can be classified according
to their (anti-)unitary symmetries and invariant bilinear forms
\cite{class-old,class-new}. 
The reason for this universal behavior is that many physical
systems and random matrix models alike can be reduced to field
theories with only Goldstone degrees of freedom. 
On general grounds, such a theory is a
nonlinear $\sigma$-model and is determined uniquely by the pattern of
symmetry breaking and convergence requirements. One can therefore
obtain the microscopic correlation functions from symmetry arguments
alone without ever referring to a random matrix model. The
construction of the nonlinear $\sigma$-models from symmetries is
standard and well known for Hermitian systems \cite{class-old,fying}.

For nonhermitian problems, which appear in open quantum systems,
e.g., in the theory of $S$-matrix fluctuations, or Euclidean QCD at
nonzero chemical potential, the situation has been investigated to a
much lesser extent. Here the formulation of $\sigma$-models based on
symmetries has been studied in detail only for the case of partition
functions given by products of determinants (i.e., {\it fermionic}
theories) \cite{KSTVZ,TV,Nish,SplitVerb2}. A major difference between
fermionic theories and bosonic or supersymmetric ones is that there
are no convergence problems for Grassmann integrals. For bosonic
nonhermitian partition functions (with inverse determinants) only
very few derivations of a $\sigma$-model from the underlying
symmetries can be found in the literature; in fact, the only works
known to us are \cite{SplitVerb2,SVbos}. In the supersymmetric case
(i.e., with both fermions and bosons) we are aware of only one model
-- the generating function of the spectral density of a Hermitian
ensemble deformed by an antihermitian ensemble -- where a
$\sigma$-model for the Goldstone degrees of freedom
\cite{sommers,sommers-rev,Efetov-direct} has been obtained. However,
that $\sigma$-model was derived by direct calculation, not by
symmetry arguments. The main objective of the present paper is to
show that a $\sigma$-model with exclusively Goldstone degrees of
freedom can be obtained from symmetry arguments alone also in
supersymmetric cases.

In this paper we study the symmetry class whose simplest
representative is a model where a complex Hermitian Gaussian random
matrix ensemble is deformed by a complex antihermitian Gaussian
random matrix ensemble. This model was introduced by Fyodorov,
Khoruzhenko and Sommers \cite{sommers} and for this reason it will be
called the FKS model. Its introduction was motivated by the study of
the distribution of resonance poles for systems with broken
time-reversal invariance \cite{S2}. More recently it was used to
describe the Hatano-Nelson model \cite{Efetov-direct} and, in its
unquenched form, QCD in three dimensions at nonzero chemical
potential \cite{Nish,GernotQCD3}.

In this paper we study the FKS model \cite{sommers} for one bosonic
flavor, the partition function of which is defined by
\begin{equation}\label{zef-1}
    Z_{-1}(z;a) = \left\langle \frac 1{\mathrm{Det}(z+H+A)}
    \right\rangle.
\end{equation}
The average is over the Gaussian probability distribution with
distribution function
\begin{equation}\label{PHA}
    P(H,A) = \mathrm{e}^{-\frac{N}{2} {\rm Tr}\, H^2 -
    \frac {N}{2a^2}{\rm Tr}\, A^2},
\end{equation}
with $H$ a Hermitian $N\times N$ matrix and $A$ an antihermitian
$N\times N$ matrix. Note that $P(H,A)$ is invariant by a unitary
change of basis, $H \mapsto g H g^{-1}$, $A \mapsto g A g^{-1}$, $g
\in \mathrm{U}(N)$. As was argued in \cite{SVbos}, there is a major
difference between fermionic and bosonic partition functions: in the
large-$N$ microscopic limit, the partition function with one
fermionic flavor does not depend on the nonhermiticity parameter,
whereas the bosonic partition function (\ref{zef-1}) does. This
behavior was found in \cite{SVbos} using the method of complex
orthogonal polynomials for a random matrix model of QCD at nonzero
chemical potential \cite{O}. Its explanation was based on the
observation that the partition function (\ref{zef-1}) needs to be
regularized. In the context of a $\sigma$-model formulation, the
technical reason is that an inverse determinant of a nonhermitian
matrix cannot be written as a Gaussian integral in general. The
regularized partition function \cite{SVbos}
\begin{equation}\label{zef-2}
    Z_{-1}(z_f^*|z,z^*;a) = \lim_{\epsilon \to 0} \left\langle
    \mathrm{Det}(z^*_f +H- A)\, \mathrm{Det}^{-1} \mat \mathrm{i}
    \epsilon & z+H+A \\ z^* + H - A & \mathrm{i}\epsilon \emat
    \right\rangle
\end{equation}
reduces to the partition function for one bosonic flavor for $z_f^*
\to z^*$. It has flavors with opposite charges resulting in a ground
state which rotates as a function of the nonhermiticity parameter.
For one fermionic flavor no regularization is necessary and the
ground state does not rotate, so that the free energy does not depend
on the nonhermiticity parameter. The regularization procedure of the
inverse determinant is known as Hermitization
\cite{janik,feinberg,Efetov-direct}.

The partition function (\ref{zef-2}) is well understood for $z_f^*\to
\infty$ in which case it is the two-flavor phase quenched bosonic
partition function. In that case, because of a complex conjugated
singularity, it diverges logarithmically with $\epsilon$ \cite{AOSV}.
In $\sigma$-model language, the singularity is due to a Goldstone
boson with a mass that vanishes as $\epsilon$ for $\epsilon \to 0$
\cite{SVbos}. Also, the partition function (\ref{zef-2}) for
$\epsilon \to 0$ acquires a Goldstone fermion of mass $z^\ast -
z_f^\ast\,$. We therefore expect the behavior
\begin{equation}\label{structure}
    Z_{-1}(z_f^*|z,z^*;a) \sim (z^* - z_f^*) \log \epsilon
    + {\cal O} (\epsilon^0).
\end{equation}
In the present paper we will derive this result from the
$\sigma$-model for the microscopic limit of (\ref{zef-2}). We will
also show that the $\mathcal{O}(\epsilon^0)$ term agrees  with the
partition function (\ref{zef-1}) evaluated by the method of complex
orthogonal polynomials.

A major issue with $\sigma$-models for nonhermitian random matrix
models is the proper choice of integration manifold. There are two
important developments which have made this choice much less ad hoc.
The first of these was the introduction of the so-called
Ingham-Siegel integral \cite{fying} as an alternative of the
Hubbard-Stratonovich transformation. This work provided a simple
explanation of the structure of the integration manifold for inverse
determinants when convergence arguments are essential. The second
development was the introduction of superbosonization which extends
the Ingham-Siegel approach to include fermions in a unified fashion
\cite{LSZ,BEKYZ}. Earlier versions of superbosonization appeared in
\cite{hack,Lehmann,schwiete0,schwiete1,guhr}, but the method was put
on a mathematically rigorous footing only in \cite{LSZ,BEKYZ}.
Although it is straightforward to obtain the correct $\sigma$-model
using the superbosonization formula of \cite{LSZ,BEKYZ}, it can be a
technically challenging task to evaluate the integrals for more than
a few degrees of freedom. The superbosonization method was applied to
nonhermitian chiral random matrix ensembles in \cite{basile}.
However, the construction of a nonlinear $\sigma$-model containing
exclusively Goldstone degrees of freedom was not given in that paper.

In this paper we will derive universal results for the symmetry class
of the FKS model relying on symmetry arguments only. This makes it
manifest that our results apply to all models in the same symmetry
class and a mass gap. Analytical results for \emph{finite} $N$ cannot
be obtained from general arguments and require a detailed
calculation. Such results will be derived for the FKS model with
Gaussian probability distributions using two independent methods, the
superbosonization method and the complex orthogonal polynomial
method. Each method has its own merits and both deserve to be
discussed. In particular, relations between partition functions with
different degrees of freedom appear naturally in the complex
orthogonal polynomial method. Using superbosonization the universal
$\sigma$-model can be recovered from the finite-$N$ results by taking
the microscopic limit and eliminating the massive modes.

Finally, let us mention that the quenched spectral density of the FKS
model has also been derived \cite{SplitVerb2} by means of the replica
limit of the Toda lattice equations \cite{SplitVerb1}. It was shown
that the quenched spectral density is the product of a fermionic and
a bosonic partition function. To derive the spectral density for one
fermionic flavor using the Toda lattice hierarchy, one needs
precisely the partition function (\ref{zef-2}) \cite{AOSV}.

In this paper we will first derive the $\sigma$-model for the
microscopic limit of (\ref{zef-2}) using symmetry arguments only
(Section II). Results for finite $N$ will be derived using the
superbosonization formula (Section III) and the complex complex
orthogonal polynomial method (Section V). In Section IV we will
recover the universal $\sigma$-model from the finite-$N$ results for
the FKS model. Concluding remarks are made in Section VI.

\section{Symmetries and supersymmetric $\sigma$-model}

In this section we will derive the universal supersymmetric
$\sigma$-model for the symmetry class of which the partition function
(\ref{zef-2}) is the simplest representative. The derivation is based
on the {\it symmetries} of (\ref{zef-2}) only, and is valid for any
other model in the same symmetry class. We will first consider the
bosonic sector with one bosonic flavor $\phi_+$ and one conjugate
bosonic flavor $\phi_-\,$. An earlier study of this sector was made
in \cite{SplitVerb2}.

\subsection{The Phase Quenched Bosonic Partition Function}
\label{sect:IIA}

The regularized phase quenched bosonic partition function can be
written as
\begin{equation}\label{pq-bos}
    Z_{\rm pq-bos}(z,z^*;a) = \left\langle \mathrm{Det}^{-1}
    \begin{pmatrix} \mathrm{i} \epsilon & z+H+A \\ z^*+H-A
    &\mathrm{i}\epsilon \end{pmatrix} \right\rangle .
\end{equation}
We will evaluate this partition function in the microscopic limit,
keeping $N \mathrm{Im} z$ and $N a^2$ fixed as $N \to \infty$. To
start the argument, we cast the inverse determinant in the form of a
Gaussian integral:
\begin{equation}\label{pq-bos1}
    Z_{\rm pq-bos}(z,z^*;a) = \int \prod_{k=1}^N d\phi_+^k
    d\phi_+^{\ast k} d\phi_-^k d\phi_-^{\ast k} \left\langle \exp\,
    \mathrm{i} \left(\begin{array}{c} \phi_+^{\ast }\,\,\phi_-^\ast
    \end{array} \right) \begin{pmatrix} \mathrm{i}\epsilon &z+H+A\\
    z^*+H-A &\mathrm{i}\epsilon\end{pmatrix}\left(\begin{array}{c}
    \phi_+ \\ \phi_- \end{array} \right) \right\rangle \;.
\end{equation}
After averaging over $H$ and $A$ the partition function can be
expressed as an integral over the positive Hermitian $2 \times 2$
matrix $Q$ of $\mathrm{U}(N)$-invariant variables
\begin{equation}
    Q = \sum_{k=1}^N  \left(\begin{array}{c} \phi_+^{\ast k}\\
    \phi_-^{\ast k} \end{array} \right) \otimes (\phi_+^k \;
    \phi_-^k) \equiv \mat \phi_+^\ast \phi_+^{\vphantom{\ast}}
    &\phi_+^\ast\phi_-^{\vphantom{\ast}} \\ \phi_-^\ast \phi_+^{
    \vphantom{\ast}}& \phi_-^\ast \phi_-^{\vphantom{\ast}}\emat .
\end{equation}
If we were dealing with fermions and compact symmetries, we could now
consider a form of `maximum flavor symmetry' using the theoretical
arguments of Peskin \cite{Peskin,KSTVZ}. A nonzero expectation value
of such a form would signal spontaneous symmetry breaking in the
thermodynamic limit. Our analysis must be somewhat different,
however, as we are facing the case of noncompact bosons. To express
the microscopic limit of the partition function (\ref{pq-bos}) in
terms of the Goldstone degrees of freedom residing in $Q$, we will
require that the resulting integration measure and Lagrangian have
the same transformation properties as the corresponding objects of
the original partition function.

The complex group $\mathrm{GL}(2)$ acts on the matrix $Q$ and the
vector variables $\phi \equiv (\phi_+ \, \phi_-)$ and $\phi^\ast =
\left( \begin{array}{c} \phi_+^\ast \\\phi_-^\ast \end{array}
\right)$ as
\begin{equation}
    \phi^\ast \mapsto g \phi^\ast \;, \quad \phi \mapsto \phi\,g^\dagger
    \;, \quad Q \mapsto g Q g^\dagger \;, \quad g \in \mathrm{GL}(2) \;.
\end{equation}
Under such transformations the integration measure $\prod_{k=1}^N
d\phi_+^k d\phi_-^k d\phi_+^{\ast k}  d\phi_-^{\ast k}$ gets
multiplied by the Jacobi determinant $\vert \mathrm{Det}\, g
\vert^{2N}$. Thus, arguing by symmetry and equating the
transformation behaviors, the corresponding measure in the
$Q$-variables is inferred to be $\mathrm{Det}^N(Q)\, dQ$ where $dQ$
denotes a $\mathrm{GL}(2)$-invariant measure for $Q$. Note also that
the trans\-formation law $Q \mapsto g Q g^\dagger$ preserves the
properties of Hermiticity and positivity of the matrix $Q$.

For $z = z^*$, $a = 0\,$, and $\epsilon \to 0$ we see that the
partition function (\ref{pq-bos}) is invariant under the subgroup $G
\subset \mathrm{GL} (2)$ of flavor transformations $Q \mapsto T Q
T^\dagger$ which preserve the Hermitian quadratic form
\begin{equation}\label{eq:herm}
    \phi_+^\ast \phi_-^{\vphantom{\ast}} + \phi_-^\ast
    \phi_+^{\vphantom{\ast}} = \mathrm{Tr}\, Q \sigma_1\;, \qquad
    \sigma_1 = \begin{pmatrix} 0 &1\\ 1 &0 \end{pmatrix} \;,
\end{equation}
of the bosonic degrees of freedom $\phi\,$. Equivalently, the
matrices $T \in G$ are subject to the condition $T^\dagger \sigma_1 T
= \sigma_1\,$. Since the Hermitian quadratic form determined by
$\sigma_1$ is of signature $(1,1)$ --- to see this, one makes a
unitary conjugation transforming $\sigma_1$ into $\sigma_3 =
\mathrm{diag}(1,-1)$ --- the symmetry group $G$ of our problem is
identified as $G = \mathrm{U}(1,1)$.

Next we search for the manifold of Goldstone degrees of freedom (or
the target space of the nonlinear $\sigma$-model) inside the space of
matrices $Q$. For that purpose, we make a temporary change of
variables from $Q$ to $X := \mathrm{i}Q \sigma_1\,$. The new matrices
$X$ satisfy $X = - \sigma_1 X^\dagger \sigma_1$ and thus lie in the
Lie algebra $\mathrm{Lie}(G)$ of $G = \mathrm{U}(1,1)$. Now, using $Q
\mapsto T Q T^\dagger $ and the relation $T^\dagger \sigma_1 =
\sigma_1 T^{-1}$ we see that $G$ acts on $X \in \mathrm{Lie}(G)$ by
the adjoint representation $X \mapsto T X T^{-1}$.

The space of $2 \times 2$ matrices $Q$ subject to the conditions $Q =
Q^\dagger > 0$ is a cone of real dimension four. Writing
\begin{equation}
    Q = \begin{pmatrix} Q_{++} &Q_{+-}\\ Q_{-+} &Q_{--} \end{pmatrix}
    = Q^\dagger \;,
\end{equation}
this cone is given by the inequalities $Q_{++} > 0\,$, $Q_{--} > 0\,
$, and $Q_{+-} Q_{-+} = | Q_{+-} |^2 < Q_{++} Q_{--}\,$. Thus our new
matrices $X = \mathrm{i} Q \sigma_1$ do not occupy the entire Lie
algebra of $G$ but lie in the cone $C^+ \subset \mathrm{Lie}(G)$
which arises as the corresponding image by the map $Q \mapsto
\mathrm{i} Q \sigma_1\,$. The cone $C^+$ may be viewed as the `space
of states' of our problem.

Let us look at the positive cone $C^+$ in a little bit of detail.
First, notice that $\mathrm{Lie}(G)$ is generated as a Lie algebra
over the real numbers by the four generators $\mathrm{i}\mathbf{1}$,
$\mathrm{i}\sigma_1\,$, $\sigma_2\,$, and $\sigma_3\,$. Let $K\subset
G$ be the maximal compact subgroup which is generated by the first
two, $\mathrm{i} \mathbf{1}$ and $\mathrm{i}\sigma_1\,$. Thus $K$ is
the group $\mathrm{U}(1) \times \mathrm{U}(1)$ of elements
$\mathrm{e}^{ \mathrm{i}(\alpha + \beta \sigma_1)}$ with $\alpha,
\beta \in [0,2\pi]$. Now consider $\mathfrak{k}^+ := C^+ \cap
\mathrm{Lie}(K)$, the intersection of the cone $C^+$ with the Lie
algebra of $K$. One may ask whether the elements $\xi \in
\mathrm{Lie}(G)$ can be conjugated into $\mathfrak{k}^+$ by the
adjoint action $\xi \mapsto T \xi T^{-1}$ of $G$. The answer is that
this is not possible in general, because $\mathrm{Lie}(G)$ is the Lie
algebra of a noncompact group. Nevertheless, the cone $C^+$
\emph{does} have the special property that each of its elements is
conjugate to some $\lambda \in \mathfrak{k}^+$ by the adjoint action
of $G$. (This follows from basic principles of linear algebra and Lie
theory and can be easily verified by direct calculation for our
simple case of $2 \times 2$ matrices.) Thus each $\xi \in C^+$ can be
presented in the `diagonalized' form
\begin{equation}\label{eq:diag}
    \xi = T \lambda T^{-1} \;, \quad T \in G \;, \quad \lambda \in
    \mathfrak{k}^+ \;.
\end{equation}
To introduce the proper mathematical language, we say that each orbit
of the adjoint $G$-action on $C^+$ hits the slice $\mathfrak{k}^+
\subset \mathrm{Lie}(K)$ at least once (actually, exactly once). It
follows from the diagonalization (\ref{eq:diag}) and the abelian
nature of $K = \mathrm{U}(1) \times \mathrm{U}(1)$ that $C^+$ has the
structure of a direct product $(G/K) \times \mathfrak {k}^+$. (More
generally, in the case of a nonabelian group $K$, the cone $C^+$ is
an associated bundle $G \times_K \mathfrak{k}^+$.) Moreover, since $K
\subset G$ is a maximal compact subgroup, the quotient $G/K$ is a
symmetric space of noncompact type. In the case at hand we get the
identification
\begin{equation}
    G / K = \mathrm{U}(1,1)/ \mathrm{U}(1) \times \mathrm{U}(1)
    = \mathrm{H}^2
\end{equation}
with a two-dimensional hyperboloid $\mathrm{H}^2$. To summarize the
present discussion: the space of states of our problem, the positive
cone $C^+$, has a decomposition (mathematically speaking, a
`fibration') $C^+ = (G/K) \times \mathfrak{k}^+$ by adjoint
$G$-orbits all of which are isomorphic to the same noncompact
symmetric space $G/K$.

Now, by some dynamical principle beyond the reach of symmetry
arguments, the system selects one of the $G$-orbits of the fibration
$C^+ = (G/K) \times \mathfrak{k}^+$ for its Goldstone manifold or
vacuum orbit. This $G$-orbit will in general be specified by an
element $\lambda \in \mathfrak{k}^+ \otimes \mathbb{C}$ of the
complexification of $\mathfrak{k}^+$. In the case under consideration
we have
\begin{equation}
    \lambda = \mathrm{i} \lambda_0 \mathbf{1} +
    \mathrm{i} \lambda_1 \sigma_1 \;,
\end{equation}
where $\lambda_0$ and $\lambda_1$ would have to be real numbers 
(with $|\lambda_0| < \lambda_1$) in
order for $\lambda$ to be in $\mathfrak{k}^+$, but in view of the
principle of steepest descent (or deformation of the integration
contour into the complex plane) we should be prepared for $\lambda_0$
and/or $\lambda_1$ to deviate from the real axis. For example, in the
case of the Gaussian ensemble (\ref{PHA}) one finds that the
$Q$-integral for $z = 0\,$, $a = 0\,$, and $\epsilon \to 0$ has a
saddle point (a maximum of the integrand) at $\lambda_0 = 0$ and
$\lambda_1 = 1\,$, with $\lambda_0$ becoming imaginary as
$\mathrm{Re}\, z$ moves away from zero. In general, $\lambda_0$ and
$\lambda_1$ take some other values. While these may be hard to
compute, we will see that the universal results emerging in the
microscopic limit do not depend on them.

Returning to our original notation, we have identified a Goldstone or
saddle-point manifold of matrices $Q$:
\begin{equation}
    Q = T Q_0 T^\dagger \;, \quad Q_0 = \lambda_1 \mathbf{1}
    + \lambda_0 \sigma_1 \;, \quad T \in G \;.
\end{equation}
From the discussion above, we know that this $G$-orbit $Q = T Q_0
T^\dagger$ is always isomorphic to the quotient $G/K$ of the
noncompact group $G$ by a maximal compact subgroup $K \subset G$.

There exist very many ways of parameterizing the $G$-orbit $Q = T Q_0
T^\dagger$. One possible choice is by a diffeomorphism $\mathrm{H}^2
\simeq \mathbb{R}^2$, exponentiating the real plane $\mathbb{R}^2$
spanned by the generators $\sigma_2$ and $\sigma_3$ as follows:
\begin{equation}\label{explit2}
    T = \mathrm{e}^{u\sigma_3 /2} \mathrm{e}^{s \sigma_2 /2}\;,\quad
    Q = T Q_0 T^\dagger = \lambda_1 \, \mathrm{e}^{u \sigma_3 / 2}
    \mathrm{e}^{s \sigma_2} \mathrm{e}^{u \sigma_3/2} + \lambda_0
    \sigma_1\;, \quad \sigma_2 = \begin{pmatrix} 0 &-\mathrm{i}\\
    \mathrm{i} &0 \end{pmatrix}\;, \quad u,s \in \mathbb{R}.
\end{equation}

Let us briefly pause to mention the following heuristic confirming
the present scenario. Suppose we were to go beyond the microscopic
limit and construct a nonlinear $\sigma$-model of spatially
fluctuating Goldstone modes with target space $G/K$. To give a
sensible definition of the functional integral of such a field
theory, we need the target space to be Riemannian. Now, for the case
of a semisimple noncompact Lie group $G$ it is a fact of differential
geometry that there is only one way to get a Riemannian manifold with
a $G$-invariant geometry: divide $G$ by a maximal compact subgroup
$K$. In contrast, the situation for fermions with compact symmetries
is very different. There, the fibration of the state space by orbits
of the symmetry group typically contains orbits of several types,
corresponding to a variety of nonisomorphic compact Riemannian
symmetric spaces. In that situation, unlike what we are facing here,
one has to appeal to a postulate of `maximum flavor symmetry'
\cite{Peskin} to select the proper type of vacuum orbit.

We are now getting ready to switch on the perturbations $a\,$,
$\epsilon\,$, and $z \not= z^\ast$ breaking $G$-symmetry. Using the
transformation law $Q \mapsto T Q \, T^\dagger$ for the Goldstone
degrees of freedom, the partition function (\ref{pq-bos1}) in the
presence of the symmetry-breaking terms remains unchanged if we
simultaneously transform
\begin{equation}\label{invA}
    \zeta \mapsto T^{\dagger \, -1} \zeta \, T^{-1} , \qquad
    {\cal A} \mapsto T^{\dagger \, -1} {\cal A}\, T^{-1} ,
\end{equation}
with
\begin{equation}
    \zeta = \mat \mathrm{i}\epsilon & \mathrm{i}\, \mathrm{Im} z \\
    - \mathrm{i}\, \mathrm{Im} z & \mathrm{i}\epsilon \emat^T \quad
    \text{and} \quad {\cal A} = \mat 0 & -\mathrm{i}a \\
    \mathrm{i} a & 0 \emat = a \sigma_2 \;.
\end{equation}
Of course the low-energy limit of the partition function
(\ref{pq-bos1}) must inherit the invariance under the transformation
(\ref{invA}).

Here, to proceed, we make the assumption that the low-energy measure
which is induced on the $G$-orbit $G/K$ of the global mode (or zero
mode) converges to the $G$-invariant measure, $d\mu(Q)$, when the
regularization parameter $\epsilon$ is taken to zero. In the case of
a compact symmetry group $G$ this assumption always holds true.
However, in the present case of a noncompact symmetry $G$ (more
precisely: a `nonamenable' symmetry $G$, see \cite{NS}), 
the $G$-invariant measure
is intrinsically unstable with respect to interactions of the
Goldstone modes. This circumstance causes a breakdown \cite{NS} of
the standard scenario of spontaneous symmetry breaking, zero mode
approximation, and universality.

Nevertheless, in the microscopic limit, i.e., for weakly interacting
Goldstone modes in a small enough volume, the said assumption does
hold true, and the integration measure on $G/K$ in the limit of $a =
0\,$, $\mathrm{Im} z = 0\,$, and $\epsilon \to 0\,$, is the
$G$-invariant measure $d\mu(Q)$. The $G$-invariance of the low-energy
partition function then forces the low-energy Lagrangian to be
$G$-invariant as well. For the mass term there exists only a single
invariant to lowest order in $\zeta:$
\begin{equation}
    {\rm Tr}(\zeta Q) = \lambda_1 \mathrm{Tr} (\zeta T T^\dagger) \;.
\end{equation}
After averaging, there are no terms linear in $a\,$. To order
$\mathcal{O}(a^2)$ there are two possible invariants:
\begin{equation}\label{inva2}
    {\rm Tr}(\mathcal{A} Q \mathcal{A} Q) \qquad {\rm and }
    \qquad {\rm Tr} (\mathcal{A} Q)\, {\rm Tr}(\mathcal{A} Q).
\end{equation}
While these invariants are independent in general, it so happens in
the present case of a single flavor that they are accidentally the
same. To verify this, one may exploit the parametrization
(\ref{explit2}) and the relation $\mathrm{e}^{u \sigma_3} \sigma_2 =
\sigma_2\, \mathrm{e}^{-u \sigma_3}$ to find the expressions
\begin{equation}\label{eq:mrz-16}
    \mathrm{Tr}(\sigma_2 Q) = 2 \lambda_1 \sinh s \;,
    \qquad \mathrm{Tr}(\sigma_2 Q\, \sigma_2 Q) =
    2\lambda_1^2 (1 + 2 \sinh^2 s) - 2 \lambda_0^2 \;,
\end{equation}
which show that $\mathrm{Tr}(\sigma_2 Q \, \sigma_2 Q) - \mathrm{Tr}
(\sigma_2 Q)\, \mathrm{Tr}(\sigma_2 Q)$ is a constant independent of
$u$ and $s$.

Thus we need only include the first invariant in the expression for
the partition function. Terms of higher order in $\zeta$ and $a^2$ do
not contribute in the microscopic limit and will not be considered
here. We also see that the unknown parameter $\lambda_0$ just adds to
the low-energy Lagrangian an inessential constant, which will not be
considered any further here (i.e., we set $\lambda_0 = 0)$. The
remaining unknown $\lambda_1$ is determined by the eigenvalue density
of the system, and we may take its value to be $\lambda_1 = 1$ by an
appropriate choice of units. Note also that $\mathrm{Det}^N(Q) =
\mathrm{Det}^N(T T^\dagger) = | \mathrm{Det}\, T |^{2N} = 1$. We thus
find that the microscopic limit of the phase quenched bosonic
partition function is given by \cite{SplitVerb2}
\begin{equation}\label{sigma}
    Z_{\rm pq-bos}(z,z^*;a) = \int d\mu(Q) \; \mathrm{e}^{\mathrm{i}
    N {\rm Tr}\, \zeta Q - \frac{1}{2} N a^2\, {\rm Tr} (Q \sigma_2
    Q\sigma_2)} \;,
\end{equation}
which is an integral over the coset space $G/K$ of matrices $Q = T
T^\dagger$ with $G$-invariant measure $d\mu(Q)$.

Now the two-hyperboloid $G/K = \mathrm{U}(1,1) / \mathrm{U}(1) \times
\mathrm{U}(1)$ is the simplest member of a certain family -- the
Hermitian symmetric spaces -- with many wonderful properties. In
particular, Hermitian symmetric spaces are K\"ahler manifolds and
come with a $G$-invariant, closed and non-degenerate two-form,
$\omega$ (the K\"ahler form). In the case at hand,
\begin{equation}
    \omega = -\mathrm{i}\,\mathrm{Tr}(\sigma_1 T^{-1} dT \wedge
    T^{-1} dT) \;,
\end{equation}
which is clearly invariant under left translations $T \mapsto g T$
corresponding to the $G$-action $Q \mapsto g\, Q \, g^\dagger$, and
also pushes down to a well-defined form on the quotient $G/K$. Using
$d^2 = 0$ and $d(T^{-1}) = - T^{-1} (dT) T^{-1}$ the expression for
$\omega$ simplifies to
\begin{equation}\label{kaehler}
    \omega = \mathrm{i}\, d\, \mathrm{Tr} (\sigma_1 T^{-1} dT) .
\end{equation}
We will shortly use this formula to compute the expression of our
$G$-invariant measure $d\mu(Q)$ in suitable coordinates.

To calculate the integral (\ref{sigma}) we use the parametrization
(\ref{explit2}), and we note that $\mathrm{Tr} (\zeta Q) = \mathrm{i}
\epsilon\, \mathrm{Tr}\,Q + \mathrm{Im}(z) \mathrm{Tr}\, \sigma_2 Q$.
From (\ref{eq:mrz-16}) we already have the expressions for
$\mathrm{Tr}\, \sigma_2 Q$ and $\mathrm{Tr} (\sigma_2 Q)^2$, and for
the remaining term in the exponent we find $\epsilon\, \mathrm{Tr} \,
Q = 2 \epsilon \cosh u \cosh s\,$. To express the measure $d\mu(Q)$
of integration we insert the parametrization (\ref{explit2}) for $T$
into (\ref{kaehler}) to obtain
\begin{equation}\label{kaehlerform}
    \omega = \mathrm{i}\, d\,\mathrm{Tr} (\sigma_1 T^{-1} dT) =
    {\textstyle{\frac{\mathrm{i}}{2}}}\, d\, \mathrm{Tr} (\sigma_1\,
    \mathrm{e}^{-s \sigma_2}\sigma_3)\wedge du = d(\sinh s)\wedge du\;.
\end{equation}
From this result we can say immediately how the measure $d\mu(Q)$
looks in the present coordinates. Indeed, since the form $\omega$ is
$G$-invariant, so is the integration measure $d(\sinh s)\, du$
corresponding to it. Because the measure $d\mu(Q)$ is determined
uniquely (up to multiplication by a constant) by $G$-invariance, we
conclude that $d\mu(Q) \propto d(\sinh s)\, du$.

Assembling terms, the phase quenched partition function (\ref{sigma})
becomes
\begin{equation}
    Z_{\rm pq-bos}(z, z^*; a) = \lim_{\epsilon \to 0+} \int_\mathbb{R}
    \mathrm{e}^{2 \mathrm{i} N \mathrm{Im}\, z \, \sinh s - N a^2
    (1 + 2 \sinh^2 s)} \left( \int_\mathbb{R} \mathrm{e}^{-2N \epsilon
    \cosh u \,\cosh s} du \right)  d(\sinh s) .
\end{equation}
The inner integral over $u$ diverges as $|\log \epsilon|$ for
$\epsilon \to 0$. The outer integral over $s$ is then a Gaussian
integral in $\sinh s$ which is easily done by completing the square.
Thus our final result \cite{SplitVerb2} for the partition function is
\begin{equation}\label{Zpq-final}
    Z_{\rm pq-bos}(z, z^*; a) = |\log \epsilon| \, \sqrt{
    \frac{\pi}{2Na^2} }\, \mathrm{e}^{- N a^2 - \frac{N}{2}
    \,\mathrm{Im}^2(z/a)} \qquad (\epsilon \to 0) .
\end{equation}

\subsection{The Partition Function for one Boson}

In this subsection we analyze the partition function (\ref{zef-2}).
To that end, we express the determinant in the numerator of
(\ref{zef-2}) as a Gaussian integral over a $\mathrm{U}(N)$
fundamental vector $\psi$ of Grassmann variables $\psi^k$. We then
combine $\psi$ with the boson flavors $\phi_\pm$ to form a
supervector $\Phi = (\phi_+ \, \phi_- \, \psi)$ with adjoint
\begin{equation}
    \Phi^\ast = \begin{pmatrix} \phi_+^\ast \\
    \phi_-^\ast \\ \bar\psi \end{pmatrix} .
\end{equation}
The low-energy effective degrees of freedom will emerge from a
supermatrix $Q$ of $\mathrm{U}(N)$-invariants,
\begin{equation}\label{supermatrix}
    Q = \Phi^\ast \Phi = \begin{pmatrix}\phi_+^\ast
    \phi_+^{\vphantom{+}} &\phi_+^\ast \phi_-^{\vphantom{+}}
    &\phi_+^\ast \psi \\ \phi_-^\ast \phi_+^{\vphantom{-}}
    &\phi_-^\ast \phi_-^{\vphantom{-}} &\phi_-^\ast \psi \\
    \bar\psi \phi_+ &\bar\psi \phi_- &\bar\psi \psi\end{pmatrix}
    \qquad (\bar\psi \psi \equiv \sum\nolimits_k \bar\psi^k
    \psi^k, \; \textrm{etc.}).
\end{equation}
Note that the boson-boson block of $Q$ is Hermitian and positive as
before. The matrix entry $Q_{ff} \equiv \bar\psi \psi$ of the
fermion-fermion block will acquire a nonzero vacuum expectation value
and is treated hence as a complex number.

Guided by the symmetries of the microscopic theory, we are now going
to identify the low-energy degrees of freedom and the structure of
the low-energy Lagrangian. On general field-theoretic grounds, we
expect the low-energy theory to be a nonlinear $\sigma$-model of
interacting Goldstone modes where the target manifold is a symmetric
space.

To see why the target space has to be symmetric --- we briefly recall
the argument here --- one may invoke Friedan's work \cite{friedan} on
the renormalization of nonlinear models and $\sigma$-models, which
shows that the quantum loop corrections to the target space metric
are given by contractions of the Riemann curvature tensor; the
one-loop correction, in particular, is given by the Ricci curvature.
(These results, while derived in the classical setting, remain valid
in the supersymmetric context.) Therefore, in a low-energy fixed
point theory the Ricci curvature of the target space must be
proportional to the metric tensor. It follows that the curvature has
to have the property of being covariantly constant which, in turn, is
the condition for a Riemannian manifold to be a symmetric space. This
result, which is fundamental for the renormalization theory of
nonlinear models and $\sigma$-models, will presently be used.

Under the most general linear transformation of the supervector
\begin{equation}
    \Phi^\ast \mapsto g_L \Phi^\ast \;, \quad \Phi \mapsto \Phi
    \, (g_R)^{-1} \;,
\end{equation}
the (a priori) superintegration form $D\Phi D\Phi^\ast = \prod_{k =
1}^N D\Phi^k D\Phi^{k\,\ast}$ transforms as
\begin{equation}\label{eq:Jacobi}
    D\Phi D\Phi^\ast \mapsto D\Phi D\Phi^\ast \mathrm{SDet}^N(g_L)
    \, \mathrm{SDet}^{-N} (g_R) \;.
\end{equation}
To match this transformation behavior, the space of composite
variables $Q$ has to be equipped with the Berezin measure (or
superintegration form) $DQ \, \mathrm{SDet}^N(Q)$ where $DQ$ by
definition is invariant under $Q \mapsto g_L Q \, (g_R)^{-1}\,$.

The Hermitian quadratic form (\ref{eq:herm}) is replaced by the
boson-fermion mixed form
\begin{equation}\label{Herm-form}
    \phi_+^\ast \phi_-^{\vphantom{\ast}} + \phi_-^\ast
    \phi_+^{\vphantom{\ast}} - \bar\psi \psi = \mathrm{STr}\,Q \Sigma_1
    \;,\qquad \Sigma_1 = \matt 0 &1 &0 \\ 1 &0 &0 \\ 0 &0 &1 \ematt .
\end{equation}
The symmetry group of this extended Hermitian form is the
pseudo-unitary Lie supergroup $G = \mathrm{U}(1,1|1)$ (a close
variant $\mathrm{U}(1,1|2)$ of which was discussed in detail in
\cite{martinbethe}).

Given the symmetry group $G = \mathrm{U}(1,1|1)$, we now ask again
about the fibration of the space of states by $G$-orbits. For that,
we temporarily switch from the supermatrices $Q$ to the related
supermatrices $X = Q \Sigma_1\,$, on which the symmetry group $G$
acts by conjugation:
\begin{equation}\label{G-action}
    Q \Sigma_1 \mapsto T (Q \Sigma_1) T^{-1} \;.
\end{equation}
We know from Section \ref{sect:IIA} that by this action the number
part of every matrix $X$ can be brought to diagonal form.

Consider first the generic case of $3 \times 3$ supermatrices $X = Q
\Sigma_1$ with three eigenvalues that all differ from one another.
The orbit of the $G$-action on such a matrix is a flag supermanifold
$\mathrm{U}(1,1|1) / \mathrm{U}(1) \times \mathrm{U}(1) \times
\mathrm{U}(1)$. Such a space is not symmetric (indeed, the Riemannian
curvature is not covariantly constant but varies) and by the
renormalizability criterion reviewed above, it can be ruled out as a
candidate for the Goldstone manifold of vacuum states.

There exists, however, the possibility for another type of $G$-orbit,
which is realized when the fermion-fermion part of $X$ becomes
degenerate with an eigenvalue of the boson-boson part. Supermatrices
$X$ on such orbits are of the form
\begin{equation}
    X = T \begin{pmatrix} \lambda_0 &\lambda_1 &0 \\ \lambda_1
    &\lambda_0 &0 \\ 0 &0 &\lambda_0 \pm \lambda_1 \end{pmatrix}
    T^{-1} , \quad T \in G \quad 
 {\rm and} \quad |\lambda_0|< \lambda_1 \;.
\end{equation}
Thus the degeneration occurs in one of two different ways: the
boson-boson part of $X$ has eigenvalues $\lambda_0 \pm \lambda_1$ and
the vacuum expectation value of $X_{ff}= \bar\psi\psi$ may hit either
one of these. In both cases our generic $G$-orbit degenerates to
\begin{equation}
    G/K \equiv \mathrm{U}(1,1|1) /
    \mathrm{U}(1) \times \mathrm{U}(1|1) \;,
\end{equation}
where $K \subset G$ is defined for $\langle \bar\psi \psi \rangle =
\lambda_0 + \lambda_1$ by the equation $k \Sigma_1 k^{-1} =
\Sigma_1\,$, and for $\langle \bar\psi \psi \rangle = \lambda_0 -
\lambda_1$ by
\begin{equation}\label{eq:Sigma1P}
    k \Sigma_1^\prime k^{-1} = \Sigma_1^\prime \;, \quad
    \Sigma_1^\prime = \begin{pmatrix} 0 &1 &0 \\ 1 &0 &0 \\
    0 &0 &-1 \end{pmatrix} \;.
\end{equation}
The quotient $G/K$ is a symmetric superspace, and thus satisfies the
renormalizability criterion, in both cases. There exists no dynamical 
or other reason (not in the microscopic limit anyway) why one of the
two $G$-orbits corresponding to the two vacuum expectation values
$\langle \bar\psi \psi \rangle = \lambda_0 \pm \lambda_1$ should be
preferred over the other. We also note that these two $G$-orbits are
\emph{disjoint}. The low-energy theory is therefore expected to be a
nonlinear $\sigma$-model with a \emph{two}-component target space,
i.e., with one connected component for each of the two vevs.

Without loss, we now simplify the discussion by setting $\lambda_0 =
0$ and $\lambda_1 = 1$ as before. The low-energy degrees of freedom
are then represented by two supermatrices $Q$ and $Q^\prime$,
\begin{equation}\label{eq:2orbits}
    Q = T \Sigma_1 T^{-1} \Sigma_1 \;, \quad
    Q^\prime = T \Sigma_1^\prime T^{-1} \Sigma_1 \;,
    \quad T \in \mathrm{U}(1,1|1) \;.
\end{equation}
Both $Q$ and $Q^\prime$ run through a symmetric superspace $G/K =
\mathrm{U}(1,1|1) / \mathrm{U}(1) \times \mathrm{U}(1|1)$, which has
the property of being \emph{Hermitian}. This fact will be of great
help in expressing the $G$-invariant integration measures $DQ$ and
$DQ^\prime$ in coordinates. Here we just note that $\mathrm{SDet}^N
(Q) = \mathrm {SDet}^N(T \Sigma_1 T^{-1} \Sigma_1) = 1$ and
\begin{equation}
    \mathrm{SDet}^N( Q^\prime) = \mathrm{SDet}^N(T \Sigma_1^\prime
    T^{-1} \Sigma_1) = (-1)^N \;.
\end{equation}

To write the result for the partition function in a concise manner,
we introduce a superscript $\sigma = \pm 1$ and let $Q^\sigma \equiv
Q$ for $\sigma = +1$ and $Q^\sigma \equiv Q^\prime$ for $\sigma =
-1$, and we denote the $G$-invariant superintegration form by
$D\mu(Q^\sigma)$. The invariance arguments of the previous section
still apply. In the microscopic limit we thus find
\begin{equation}\label{sigma3}
    Z_{-1}(z_f^*|z,z^*;a) = \sum_{\sigma = \pm 1}
    \sigma^N \int D\mu(Q^\sigma) \; \mathrm{e}^{\mathrm{i}N
    \mathrm{STr} \,\zeta Q^\sigma -\frac{1}{2} N a^2 \,
    \mathrm{STr} \, Q^\sigma \Sigma_2 Q^\sigma \Sigma_2} ,
\end{equation}
where the mass matrix $\zeta$ and the extended Pauli matrix
$\Sigma_2$ are now given by
\begin{equation}\label{zeta}
    \zeta = \matt \mathrm{i}\epsilon &z^\ast &0\\ z &\mathrm{i}
    \epsilon &0\\ 0 &0 &z_f^* \ematt, \qquad \Sigma_2 = \matt 0
    &-\mathrm{i} &0\\ \mathrm{i} &0 &0\\ 0 &0 &-\mathrm{i} \ematt .
\end{equation}

To summarize, we have expressed the microscopic limit of $Z_{-1}
(z_f^*|z,z^*;a)$ as an integral over Goldstone degrees of freedom
only. (Of course, the overall normalization factor cannot be fixed by
the arguments in this section.)
%
%
%
In the next section we will rederive the result (\ref{sigma3}) using
superbosonization.

The calculation of the integral (\ref{sigma3}) requires an explicit
parametrization of the Goldstone degrees of freedom. For this purpose
we choose a fermion-boson factorized generalization of the
parametrization (\ref{explit2}):
\begin{equation}\label{eq:new-param}
    T = T_f \, T_b \;, \quad T_f = \matt 1 &\mathbf{0}\\ \mathbf{0}
    &V \emat \;, \quad V = \mat 1 &\alpha\\ \beta &1 \emat \;, \quad
    T_b = \mat W &\mathbf{0}\\ \mathbf{0} &1 \emat \;, \quad
    W = \mathrm{e}^{u \sigma_3 / 2} \mathrm{e}^{s \sigma_2 / 2} \;.
\end{equation}
To motivate this choice, let us observe that
\begin{equation}
    \Sigma_3 := - \mathrm{i} \Sigma_1 \Sigma_2 =
    \matt 1 &0 &0\\ 0 &-1 &0\\ 0 &0 &-1 \ematt
\end{equation}
commutes with $T_f\,$. Also, let $\Sigma_1^\sigma := \Sigma_1$ for
$\sigma = +1$ and $\Sigma_1^\sigma := \Sigma_1^\prime$ for $\sigma =
-1$, so that $Q^\sigma \Sigma_1 = T \Sigma_1^\sigma T^{-1}$ in both
cases. Note $(\Sigma_1^\sigma)^2 = \mathbf{1}$. Looking at $Q^\sigma
\Sigma_2 = \mathrm{i} T_f T_b \Sigma_1^\sigma T_b^{-1} T_f^{-1}
\Sigma_3$ we see that the expression for $\mathrm{STr} (Q^\sigma
\Sigma_2)^2$ remains unchanged from (\ref{eq:mrz-16}) but for the
addition of a trivial constant:
\begin{equation}\label{eq:mrz-36}
    \mathrm{STr} (Q^\sigma \Sigma_2)^2 = 3 + 4 \sinh^2 s \;.
\end{equation}
The mass term has the expression
\begin{equation}\label{eq:massterm}
    \mathrm{STr}\, \zeta Q^\sigma = 2\mathrm{i} \epsilon \cosh s
    \cosh u - \sigma z_f^\ast - \mathrm{i}(z - z^\ast) \sinh s +
    \alpha\beta\big((z^\ast-z_f^\ast)(\mathrm{i}\sinh s - \sigma)
    + \mathrm{i} \epsilon \, \mathrm{e}^u \cosh s \big) \;.
\end{equation}

To express $D\mu(Q^\sigma)$ in coordinates we recall a few facts from
(super-)geometry. If $x^1, \ldots, x^p$ and $\xi^1, \ldots, \xi^q$
form a system of commuting and anti-commuting local coordinates for a
Riemannian supermanifold $M$ with metric tensor
\begin{equation}\label{eq:g-in-coords}
    g = A_{ij}\, dx^i dx^j + B_{ik}\, dx^i d\xi^k + C_{lj} \,
    d\xi^l dx^j + D_{kl}\, d\xi^k d\xi^l \;,
\end{equation}
where $A_{ij} = A_{ji}$ and $D_{kl} = - D_{lk}$ are even and $B_{ik}
= C_{ki}$ are odd functions of the $\xi$'s, then the associated
superintegration form is
\begin{equation}\label{eq:Berezin}
    \Omega = dx^1 \cdots dx^p \, \frac{\partial}{\partial \xi^1}
    \cdots \frac{\partial}{\partial \xi^q}\,\circ \mathrm{SDet}^{1/2}
    \begin{pmatrix} A &B\\ C &D \end{pmatrix} \;.
\end{equation}
Now our supermanifolds $M = \mathrm{U}(1,1|1) / \mathrm{U}(1) \times
\mathrm{U}(1|1)$ carry the $G$-invariant metric
\begin{equation}\label{eq:metric}
    g = 2\, \mathrm{STr}\, (T^{-1} dT)_\mathfrak{p}^2 \;,
\end{equation}
where the subscript $\mathfrak{p}$ signifies projection on the
tangent space of $M$ at the origin; i.e., $A_\mathfrak {p}$ is the
component of $A$ that anti-commutes with $\Sigma_1^\sigma$:
\begin{equation}
    A_\mathfrak{p} := {\textstyle{\frac{1}{2}}}
    \left( A - \Sigma_1^\sigma A \Sigma_1^\sigma \right) \;.
\end{equation}
Using (\ref{eq:metric}) in conjunction with (\ref{eq:g-in-coords},
\ref{eq:Berezin}) we can express $\Omega \equiv D\mu(Q^\sigma)$ in
any coordinate system of our choice.

However, in the present situation there exists a better way of doing
this calculation. Being a Hermitian symmetric superspace, $M \equiv
M^\sigma$ (for $\sigma = \pm 1$) comes with a tensor field $J$ called
a complex structure. This means that $J^2 = - \mathbf{1}$ and $J$ is
an isometry of the metric, i.e., $g(Ju,Jv) = g(u,v)$ for any two
tangent vector fields $u,v$. In our case the induced action of $J$ on
the one-form $(T^{-1}dT)_\mathfrak{p}$ is given by
\begin{equation}
    (T^{-1}dT)_\mathfrak{p} \mapsto
    \mathrm{i} \Sigma_1^\sigma (T^{-1} dT)_\mathfrak{p} =
    - \mathrm{i} (T^{-1} dT)_\mathfrak{p} \Sigma_1^\sigma \;.
\end{equation}
Given the (K\"ahler) metric $g$ and the complex structure $J$ one
defines a two-form $\omega$ (the K\"ahler form) by the equation
$\omega(u,v) = g(u,Jv)$. In the present case we find
\begin{equation}\label{eq:kaehler-s}
    \omega \equiv \omega^\sigma =
    - \mathrm{i}\, \mathrm{STr} \big( \Sigma_1^\sigma
    (T^{-1} dT)_\mathfrak{p} \wedge (T^{-1} dT)_\mathfrak{p} \big)
    = \mathrm{i}\, d \, \mathrm{STr} (\Sigma_1^\sigma T^{-1} dT) \;.
\end{equation}
For our purposes, the form $\omega$ is a useful object to introduce
because $\omega$ is easier to express than the metric $g$ and yet
carries enough information to construct the Berezin measure $\Omega =
D\mu(Q^\sigma)$. In fact, if $\omega$ is expressed in coordinates as
\begin{equation}
    \omega = {\textstyle{\frac{1}{2}}} \left( \tilde{A}_{ij}\,
    dx^i \wedge dx^j + \tilde{B}_{ik}\, dx^i \wedge d\xi^k +
    \tilde{C}_{lj} \, d\xi^l \wedge dx^j + \tilde{D}_{kl}\,
    d\xi^k \wedge d\xi^l \right) \;,
\end{equation}
where $\tilde{A}_{ij} = - \tilde{A}_{ji}\,$, $\tilde{D}_{kl} =
\tilde{D}_{lk}\,$, and $\tilde{B}_{ik} = - \tilde{C}_{ki}$ (due to
skewness of the wedge product), then we have
\begin{equation}\label{eq:sameSDet}
    \mathrm{SDet} \begin{pmatrix} A &B\\ C &D \end{pmatrix}
    = \mathrm{SDet} \begin{pmatrix} \tilde{A} &\tilde{B}\\
    \tilde{C} &\tilde{D} \end{pmatrix}
\end{equation}
as a consequence of the properties of the complex structure $J$
relating the metric $g$ with the K\"ahler form $\omega$.

We are now in a position to compute $D\mu(Q^\sigma)$ with ease.
Inserting the parametrization (\ref{eq:new-param}) into
(\ref{eq:kaehler-s}) we obtain
\begin{equation}
    \omega^\sigma = \mathrm{i}\, d\, \mathrm{STr}
    (\Sigma_1^\sigma T_b^{-1} dT_b) + \mathrm{i}\,
    d\, \mathrm{STr} (T_b \Sigma_1^\sigma T_b^{-1} T_f^{-1} dT_f)\;.
\end{equation}
The term $\mathrm{i}\, d\, \mathrm{STr} (\Sigma_1^\sigma T_b^{-1}
dT_b) = \mathrm{i}\, d\, \mathrm{Tr} (\sigma_1 W^{-1} dW) = d(\sinh
s) \wedge du$ was already computed in (\ref{kaehlerform}). The new
term is
\begin{equation}
    \mathrm{i}\, d\,\mathrm{STr} (T_b \Sigma_1^\sigma T_b^{-1}T_f^{-1}
    dT_f) = d (\sinh(s)\, \alpha\, d\beta + \mathrm{i}\sigma \beta
    d\alpha ) = (\mathrm{i}\sigma + \sinh s)\, d\alpha \wedge d\beta +
    \alpha\, d(\sinh s) \wedge d\beta \;.
\end{equation}
The last summand makes no contribution to the superdeterminant of the
metric tensor (since $\alpha^2 = 0$) and therefore can be dropped for
the purpose of constructing $D\mu(Q^\sigma)$. The term proportional
to $d\alpha \wedge d\beta = d\beta \wedge d\alpha$ contributes the
reciprocal of the analytic square root of $- (\mathrm{i} \sigma +
\sinh s)^2$. Thus from Eqs.\ (\ref{eq:sameSDet}) and
(\ref{eq:Berezin}) we have
\begin{equation}
    D\mu(Q^\sigma) = \frac{d(\sinh s)\, du}
    {1 - \mathrm{i}\sigma \sinh s}\, d\alpha\, d\beta \;.
\end{equation}
Please be advised that the symbol $d\alpha$ in this expression means
the derivative $d\alpha \equiv \partial / \partial \alpha\,$,
although its meaning in the previous equation was that of a
differential. Thus we are using the same symbol $d\alpha$ for two
very different objects.

\section{Microscopic limit of the partition function}

In this section we evaluate the partition function (\ref{sigma3}) in
two different limits. First, we extract the contribution that
diverges as $\log \epsilon$ for $z_f^* \ne z^*$, and second, we
compute the result for $z_f^* = z^*$ which is regular for $\epsilon
\to 0\,$.

\subsection{Contribution of order $\log \epsilon$}

Using (\ref{eq:massterm}) for the mass term and the expression
(\ref{eq:mrz-36}) for $\mathrm{STr} (Q \Sigma_2)^2$, the partition
function (\ref{sigma3}) becomes
\begin{eqnarray}
    Z_{-1}(z_f^*|z,z^*;a) &\sim& \mathrm{i}^{-N} \mathrm{e}^{
    -\frac{3}{2} N a^2} \sum_{\sigma = \pm 1} \sigma^N
    \mathrm{e}^{-\mathrm{i}N \sigma z_f^\ast} \int
    \frac{d(\sinh s)\,du}{1 - \mathrm{i}\sigma \sinh s} \,
    d\alpha \, d\beta  \nn \\ &\times& \mathrm{e}^{-2N\epsilon
    \cosh s \cosh u + N (z - z^*)\sinh s - 2N a^2 \sinh^2 s
    +\mathrm{i}N \alpha \beta ((z^*-z_f^*)(\mathrm{i}\sinh s
    - \sigma) + \mathrm{i}\epsilon\, \mathrm{e}^u \cosh s)}
    \label{eq:mrz-52}.
\end{eqnarray}
Here we display only the factors of alternating phase; the full
overall normalization factor will be inserted below.

The integral over the Grassmann variables $\alpha, \beta$ yields a
factor
\begin{equation}
    N \epsilon\, \mathrm{e}^u \cosh s + N \mathrm{i}
    \sigma (z^*-z_f^*)(1- \mathrm{i}\sigma \sinh s) \;.
\end{equation}
Since the rest of the integrand is even in $u$ we may replace
$\mathrm{e}^u$ in this expression by $\cosh u\,$. The $u$-integral
with the resulting term $N\epsilon\, \cosh u \cosh s$ is finite in
the limit $\epsilon \to 0$ (see next subsection). Therefore we may
drop this term here, as we are after the singular contribution
$\propto \log \epsilon\,$. The $u$-integral over the remaining term
has the asymptotics
\begin{equation}
    \int du \, \mathrm{e}^{-2N \epsilon \cosh s \cosh u} =
    2\, | \log \epsilon | + \mathcal{O}(\epsilon^0) \;.
\end{equation}
Doing finally the $s$-integral by completing the square we obtain the
leading term
\begin{equation}\label{Zm1SUSY}
    Z_{-1} (z_f^*|z,z^*; a) = c_N |\log \epsilon| \, a^{-1} (z^\ast
    - z_f^\ast)\, \mathrm{e}^{- \frac{N}{2}\, \mathrm{Im}^2 (z/a) -
    \frac{3}{2} N a^2}\sin(N z_f^*+ N\pi/2) +\mathcal{O}(\epsilon^0)\;.
\end{equation}
The normalization constant $c_N$ is found by keeping track of all
constants in the calculations above. Using the formula $\mathrm{vol}
\, \mathrm{U}(N) / \mathrm{vol}\, \mathrm{U}(N-1) = (2\pi)^N /
(N-1)!$ and Stirling's approximation for the factorial, we find
\begin{equation}
    c_N = \mathrm{e}^{N/2} (N/\pi)^{1/2} \;.
\end{equation}
This result (\ref{Zm1SUSY}) will be verified by taking the
microscopic limit of the exact finite-$N$ results that will derived
in Section \ref{sec:super} by means of superbosonization and in
Section \ref{sec:OP} using the method of complex orthogonal
polynomials.

\subsection{Contribution of order $\epsilon^0$}

For $z_f^* = z^*$ the logarithmic singularity $| \log \epsilon |$
vanishes. Indeed, doing the Grassmann integrals over $\alpha, \beta$
and the $u$-integral and then sending $\epsilon \to 0$ we obtain a
finite limit
\begin{equation}
    \lim_{\epsilon \to 0+} \epsilon N \cosh s \int_\mathbb{R}
    du\; \mathrm{e}^{u-2\epsilon N \cosh s \cosh u} = 1\;.
\end{equation}
With the substitution $q \equiv \sinh s$ the expression
(\ref{eq:mrz-52}) for the partition function now becomes
\begin{equation}\label{super}
    Z_{-1}(z^*|z,z^*;a) \sim \mathrm{e}^{-\frac{3}{2} N a^2}
    \sum_{\sigma = \pm 1} \mathrm{e}^{-\mathrm{i}\sigma N
    (z^\ast + \pi/2)} \int \frac{dq}{1 - \mathrm{i}\sigma q}
    \, \mathrm{e}^{N (z - z^*)\, q - 2N a^2 q^2} \;.
\end{equation}
Introducing an auxiliary integration by $(1 - \mathrm{i} \sigma
q)^{-1} = \int_0^\infty \mathrm{e}^{-t(1 - \mathrm{i}\sigma q)} dt$
we can do the Gaussian integral over $q$ by completing the square.
The result of this step is immediately expressed in terms of the
complementary error function:
\begin{equation}
    \int \frac{dq}{1 - \mathrm{i}\sigma q}\, \mathrm{e}^{N
    (z - z^*)\,q - 2N a^2 q^2} = \pi\, \mathrm{e}^{2N (a^2
    + \sigma \,\mathrm{Im} z)}\, \mathrm{erfc} \left( a\sqrt{2N}+
    \frac{\sigma N \mathrm{Im}\,z}{a \sqrt{2N}} \right) \;,
\end{equation}
which is defined by $\mathrm{erfc}(x) = 1 - \mathrm{erf}(x) = (2
/\sqrt{\pi}\,) \int_x^{\infty} \mathrm{e}^{-t^2} dt\,$. Our final
result reads
\begin{equation}\label{Z1bosSUSY}
    Z_{-1}(z^*|z,z^*;a) = \mathrm{e}^{\frac{1}{2} N(1+a^2)}\,
    2^{-3/2} \sum_{\sigma = \pm 1} \mathrm{e}^{-\mathrm{i}\sigma
    N (z+\pi/2)}\, \mathrm{erfc} \left(a\sqrt{2N} + \frac{\sigma
    N \mathrm{Im}\, z}{a\sqrt{2N}} \right) \;.
\end{equation}
This expression agrees with the result obtained from the microscopic
limit of the finite-$N$ results that will be derived by means of
superbosonization in the next section, and by using the Cauchy
transform of orthogonal polynomials in Section \ref{sec:OP}.

\section{Superbosonization}\label{sec:super}

As discussed in the introduction, symmetry arguments alone are not
sufficient to derive results for finite $N$. This requires an exact
evaluation of the partition function which, in this section, is
achieved by the method of superbosonization. In the next section we
will obtain exact finite-$N$ results by means of complex orthogonal
polynomials.

The superbosonization method was introduced to address problems with
nongaussian disorder
\cite{hack,Lehmann,schwiete0,schwiete1,guhr,LSZ,BEKYZ}. Its main idea
is to reduce an integral with symmetries to a lower-dimensional
integral. To give a simple example illustrating this general idea,
consider a function $f$ of complex variables $z^1, \ldots, z^N$. If
$f$ depends only on $x = \sum_{k=1}^N |z^k|^2$ then the integral of
$f$ over a $\mathrm{U}(N)$-invariant domain in $\mathbb{C}^N$ can be
reduced to an integral over just $x$. Similarly, the Grassmann
integral of a function $f(\sum_{k=1}^N \bar\psi^k \psi^k)$ of
anti-commuting variables $\psi^k$ and $\bar\psi^k$ is known
\cite{kawamoto-smit} to be expressible as an integral of $f(y)$ over
$y \in \mathrm{U}(1)$.

Based on results from invariant theory, superbosonization extends
this reduction idea to the general case of invariant functions of
supervectors. In the bosonic sector, the method is equivalent to the
one introduced in \cite{fying}. However, in \cite{fying} the
fermionic degrees of freedom were bosonized in the usual way by means
of a Hubbard-Stratonovich transformation (we will refer to this
procedure as the hybrid method), whereas in the superbosonization
approach the fermionic and bosonic variables are treated on equal
footing. From our perspective, a major advantage of the
superbosonization method is that the integration measure is given by
a general formula which can be easily applied to a specific case such
as the bosonic partition function considered in this paper. In order
to execute the integrals, it is essential that the parametrization be
chosen judiciously.

The present partition function has also been worked out in a
straightforward way using the hybrid method of \cite{fying}. That
calculation is not more complicated than the superbosonization
method, but since it does not provide us with any additional insights
we will not discuss the hybrid method any further.

The starting point for superbosonization of the regularized FKS
partition function (\ref{zef-2}) is the representation of the inverse
determinant as
\begin{equation}\label{e42}
    \mathrm{Det}^{-1} \mat \mathrm{i}\epsilon & z+H+A \\
    z^* + H - A &\mathrm{i}\epsilon \emat = \int d\phi\,
    d\phi^\ast \, \mathrm{e}^{-\epsilon (\phi_+^{* k}
    \phi_+^k + \phi_-^{* k} \phi_-^k)+ \mathrm{i} \phi_+^{* k}
    (z\delta_{kl} + H_{kl} + A_{kl})\phi_-^l + \mathrm{i}
    \phi_-^{* k}(z^*\delta_{kl} + H_{kl} - A_{kl}) \phi_+^l}
\end{equation}
and the fermion determinant as
\begin{equation}\label{e43}
    \mathrm{Det} (z^*_f+H-A) = \int d\psi\, d\bar\psi \;
    \mathrm{e}^{- \mathrm{i}\bar\psi^k (z^*_f \delta_{kl}
    + H_{kl} - A_{kl}) \psi^l}.
\end{equation}
Taking the average over the Gaussian distribution (\ref{PHA}) of $H$
and $A$ we obtain
\begin{eqnarray}
    \left\langle \mathrm{e}^{\mathrm{i} H_{kl} (\phi_+^{* k}
    \phi_-^l + \phi_-^{\ast k} \phi_+^l - \bar\psi^k \psi^l)}
    \right\rangle_H &=& \mathrm{e}^{- (1/2N)\, \mathrm{STr}\, Q
    \Sigma_1 Q \Sigma_1} \;, \\
    \left\langle \mathrm{e}^{\mathrm{i} A_{kl} (\phi_+^{* k}
    \phi_-^l - \phi_-^{\ast k} \phi_+^l + \bar\psi^k \psi^l)}
    \right\rangle_A &=& \mathrm{e}^{-(a^2/2N)\, \mathrm{STr}\, Q
    \Sigma_2 Q \Sigma_2} \;,
\end{eqnarray}
where $Q$ is the supermatrix (\ref{supermatrix}) of $\mathrm{U}
(N)$-invariant bilinears, and the matrices $\Sigma_1$ and $\Sigma_2$
were defined in (\ref{Herm-form}) and (\ref{zeta}). The quadratic
terms in the exponents of (\ref{e42}) and (\ref{e43}) combine to make
up the mass term:
\begin{equation}
    \mathrm{e}^{- \epsilon \phi_+^{\ast k} \phi_+^k - \epsilon
    \phi_-^{\ast k} \phi_-^k + \mathrm{i}z \phi_+^{\ast k}
    \phi_-^k + \mathrm{i}z^\ast \phi_-^{\ast k} \phi_+^k
    - \mathrm{i}z_f^\ast \bar\psi^k \psi^k} =
    \mathrm{e}^{\mathrm{i}\, \mathrm{STr}\, \zeta Q} \;.
\end{equation}

The method of superbosonization allows us now to introduce the matrix
elements of $Q$ directly as the new variables of integration. Using a
formula proved in \cite{LSZ} the partition function (after rescaling
$Q \to NQ$) reduces to
\begin{equation}\label{Z-supbos}
    Z_{-1}(z_f^*|z,z^*;a) = (2\pi \mathrm{i})^{-N} \frac{\mathrm{vol}
    \, \mathrm{U}(N)}{ \mathrm{vol}\, \mathrm{U}(N-1)} \int DQ \;
    \mathrm{SDet}^N (N Q)\; \mathrm{e}^{\mathrm{i} N \mathrm{STr} \,
    \zeta Q -(N/2)\, \mathrm{STr}\, Q\Sigma_1 Q\Sigma_1 - (N a^2 / 2)
    \, \mathrm{STr}\, Q\Sigma_2 Q \Sigma_2} \;.
\end{equation}
By the superbosonization step of passing to (\ref{Z-supbos}), the
precise meaning of $Q$ has been transformed: $Q$ is now the
supermatrix
\begin{equation}
    Q = \begin{pmatrix} Q_{bb} &Q_{bf}\\ Q_{fb} &Q_{ff} \end{pmatrix}
    \equiv \begin{pmatrix} X_{++} &X_{+-} &\xi_+ \\ X_{-+} &X_{--}
    &\xi_- \\ \eta_+ &\eta_- &y \end{pmatrix} \;,
\end{equation}
where the boson-boson block $Q_{bb} \equiv X$ is a positive Hermitian
matrix, $Q_{ff} \equiv y \in \mathrm{U}(1)$ is a unitary number, and
the components of $Q_{fb} \equiv \eta = (\eta_+ \, \eta_-)$ and
$Q_{bf} \equiv \xi = \begin{pmatrix} \xi_+ \\ \xi_- \end{pmatrix}$
are Grassmann variables. The Berezin measure is \cite{LSZ,BEKYZ}
\begin{equation}\label{supermeasure}
    DQ = (4\pi^2 \mathrm{i})^{-1} d^4X dy \, d \xi_+ d\eta_+ d\xi_-
    d\eta_- \; \mathrm{SDet}^{-1}(Q)\;, \qquad d^4X \propto dX_{++}\,
    dX_{--}\, dX_{+-}\, dX_{-+}\;.
\end{equation}
Following the conventions of \cite{LSZ} we normalize the flat measure
$d^4 X$ so that $\lim_{\delta \to 0} \delta^{-2} \int_{X
> 0} \mathrm{e}^{- (\pi/\delta)\, \mathrm{Tr}\, (X-\mathbf{1})^2}
d^4 X = 1$. Note that $DQ$ is scale-invariant; it is also invariant
under the transformation $Q \Sigma_1 \mapsto Q \Sigma_1^\prime$ for
$\Sigma_1^\prime$ given in (\ref{eq:Sigma1P}).

The expression (\ref{Z-supbos}) for $Z_{-1}$ is suitable for
saddle-point analysis in the limit $N \to \infty$. Since we are
considering the microscopic limit where $N \zeta$ and $N a^2$ are
held fixed as $N$ goes to infinity, the symmetry-breaking terms are
subleading in $1/N$ and can be temporarily neglected for the purpose
of finding the saddle-point manifold. If we set $\zeta = 0$ and $a^2
= 0\,$, the logarithm of the integrand becomes
\begin{equation}
    N \mathrm{STr}\, \log Q - \frac{N}{2}
    \mathrm{STr}\, Q \Sigma_1 Q \Sigma_1 \;,
\end{equation}
variation of which gives the saddle-point equation $Q^{-1} = \Sigma_1
Q \Sigma_1\,$. The solutions of this equation form the two disjoint
supermanifolds $Q = T \Sigma_1 T^{-1} \Sigma_1$ and $Q = T
\Sigma_1^\prime T^{-1} \Sigma_1$ which were described in
(\ref{eq:2orbits}). Note that the signs of the solution in the
boson-boson sector are fixed by the condition $Q_{bb} > 0\,$.

Next we perform the integration over the massive modes in the
large-$N$ limit. To handle both saddle-point manifolds at once, we
recall our notation $\Sigma_1^\sigma \equiv \Sigma_1$ for $\sigma =
+1$ and $\Sigma_1^\sigma \equiv \Sigma_1^\prime$ for $\sigma = -1$,
and we set
\begin{equation}
    Q = T \mathrm{e}^P \Sigma_1^\sigma T^{-1} \Sigma_1 \;,
\end{equation}
where the matrix $P$ parameterizes the massive modes. By the very
definition of what it means to be a massive mode, $P$ commutes with
$\Sigma_1^\sigma$ in both cases. To do the integral over $P$ to
leading order in $1/N$ we may put $P$ equal to zero in the
symmetry-breaking terms with parameters $a^2$ and $\zeta$ (we remind
the reader that both $a^2$ and $\zeta$ are of order $1/N$). Thus we
need to integrate
\begin{equation}
    \mathrm{SDet}^N(Q)\, \mathrm{e}^{-(N/2)\,\mathrm{STr}\,
    (Q\Sigma_1)^2} = \sigma^N \mathrm{e}^{N\, \mathrm{STr}\,
    P - (N/2)\, \mathrm{STr}\, \mathrm{e}^{2P}} = \sigma^N
    \mathrm{e}^{-(N/2) - 2N\,\mathrm{STr}\, P^2 + \ldots} .
\end{equation}
We see that the fluctuations of the massive modes are of the order $P
\sim 1 / \sqrt{N}$. Because of the smallness of these fluctuations we
may replace the nonlinear Berezin measure for the $P$-variables by
the \emph{flat} Berezin measure $D_0 P$ (i.e., the product of
differentials for the commuting variables and derivatives for the
anti-commuting variables). Thus we have $DQ \simeq D_0 P\,
D\mu(Q^\sigma)$ where $Q^\sigma = T \Sigma_1^\sigma T^{-1} \Sigma_1$
and $D\mu(Q^\sigma)$ is the Berezin measure which is invariant under
the transformation $Q \Sigma_1 \mapsto T Q \Sigma_1 T^{-1}$. The
integral over the massive modes $P$ then is a simple Gaussian
integral $\int D_0 P \, \mathrm{e}^{-2N \mathrm{STr}\, P^2}$. Doing
it we immediately arrive at the result (\ref{sigma3}) of the previous
section. Moreover, we are now in principle able to determine the
precise normalization constant. We will insert the correct overall
normalization when evaluating the partition function below.

\section{Exact Calculation using Superbosonization}

We now use the result (\ref{Z-supbos}) from superbosonization to
derive an exact expression for finite $N$. To that end we start from
the formula for the superdeterminant,
\begin{equation}\label{eq:SDet}
    \mathrm{SDet}\, Q = \mathrm{SDet}\begin{pmatrix} X &\xi\\ \eta
    &y \end{pmatrix} = \frac{\mathrm{Det}(X)}{y-\eta X^{-1}\xi}\;,
\end{equation}
where $\eta X^{-1} \xi$ means the scalar which is obtained by
sandwiching the matrix $X^{-1}$ between the row vector $\eta$ and the
column vector $\xi\,$. In view of Eq.\ (\ref{eq:SDet}), the factor
$\mathrm{SDet}^N(Q)$ of the integrand of (\ref{Z-supbos}) is much
simplified by making a shift $y \to y + \eta X^{-1} \xi\,$. Such a
shift leaves the integral over $y \in \mathrm{U}(1)$ invariant:
$\oint_{\mathrm{U}(1)} f(y)\, dy = \oint_{\mathrm{U} (1)} f(y + \eta
X^{-1} \xi)\, dy\,$. After this shift, our integrand depends on the
anti-commuting variables only through the following factor:
\begin{equation}
    \Phi = \mathrm{e}^{N \eta\,( A + (1-a^2) y \,) X^{-1} \xi +
    \frac{N}{2} (1-a^2) (\eta X^{-1} \xi)^2} \;, \qquad A \equiv A(X)
    = \sigma_1 X - \mathrm{i}a^2 \sigma_2 X - \mathrm{i}z_f^\ast\;.
\end{equation}
Using the relation $\frac{1}{2}(\eta X^{-1}\xi)^2 = \mathrm{Det}^{-1}
(X)\, \eta_+ \xi_+ \eta_- \xi_-$ we now carry out the integral over
the anti-commuting variables to obtain $\int d\xi_+ d\eta_+ d\xi_-
d\eta_- \, \Phi = \mathrm{Det}^{-1}(X)\, F(X,y)$ where
\begin{equation}
    F(X,y) = N (1-a^2) + N^2 \left(\mathrm{Det} \,A(X) + (1-a^2)
    \,y\, \mathrm{Tr}\,A(X) + (1-a^2)^2 y^2 \right)\;.
\end{equation}
We insert this into the integral representation (\ref{Z-supbos}) of
the partition function to get
\begin{eqnarray}
    Z_{-1}(z_f^\ast|z,z^\ast;a) &=& \frac{\mathrm{i}^{-N-1} N^N}
    {4\pi^2 (N-1)!} \int_{X > 0} d^4X \; \mathrm{Det}^{N-2}(X)
    \oint_{\mathrm{U}(1)} dy \; y^{-N+1}\, F(X,y) \nonumber \\
    &\times& \mathrm{e}^{- \epsilon N \, \mathrm{Tr}\,X +
    \mathrm{i}N \mathrm{Tr}\, (X \sigma_1 \mathrm{Re}\, z +
    X \sigma_2\, \mathrm{Im}\, z) - \mathrm{i}N z_f^\ast y -
    \frac{N}{2} \mathrm{Tr}\, (X \sigma_1 X \sigma_1 + a^2 X
    \sigma_2 X \sigma_2) + \frac{N}{2}(1-a^2) \, y^2} \;.
\end{eqnarray}

Next we do the $y$-integral. For this purpose we introduce the
$a$-dependent functions
\begin{equation}\label{eq:tilde-h}
    \tilde{h}_{N,\,k}(\tau) := \frac{(1-a^2)^k}{2\pi}
    \oint_{\mathrm{U}(1)} dy\; (\mathrm{i}y)^{-N+1+k}\,
    \mathrm{e}^{-\mathrm{i} \tau y + \frac{N}{2}(1-a^2) y^2} \;,
\end{equation}
which will be shown presently to be scaled Hermite polynomials. With
this definition, what remains to be done is an integral over the
positive Hermitian $2 \times 2$ matrices $X:$
\begin{eqnarray}
    Z_{-1} &=& \frac{N^{N+2}}{2\pi (N-1)!} \int_{X > 0} d^4X\;
    \mathrm{e}^{-\epsilon N \mathrm{Tr}\,X + \mathrm{i}N \mathrm{Tr}
    \,(X \sigma_1 \mathrm{Re}\, z + X \sigma_2\,\mathrm{Im}\, z) -
    \frac{N}{2}\mathrm{Tr}\,(X\sigma_1 X \sigma_1 + a^2 X \sigma_2 X
    \sigma_2)} \nonumber \\&\times& \mathrm{Det}^{N-2}(X) \left((
    \tilde{h}_{N,2}(N z_f^\ast) + \mathrm{i}\, \mathrm{Tr}\, A(X)\,
    \tilde{h}_{N,1}(N z_f^\ast) - (\mathrm{Det}\,A(X)+N^{-1}(1-a^2))
    \,\tilde{h}_{N,0} (N z_f^\ast) \right) \;. \label{eq:partfunc}
\end{eqnarray}

To compute the $X$-integral one may use the parametrization
\begin{equation}
    X = \begin{pmatrix} \mathrm{e}^u \sqrt{p + v^2 + w^2}
    &v - \mathrm{i} w \\ v + \mathrm{i} w &\mathrm{e}^{-u} \sqrt{
    p + v^2 + w^2} \end{pmatrix} \qquad (u,v,w \in \mathbb{R} \;,
    \; p \in \mathbb{R}_+ )\;.
\end{equation}
The integration measure in these coordinates is expressed by
\begin{equation}
    d^4 X = 2\,dp\,du\,dv\,dw\;,
\end{equation}
and some traces appearing in the exponent of the integrand are
\begin{equation}\label{eq:traces}
    {\textstyle{\frac{1}{2}}} \mathrm{Tr} (X \sigma_1 X \sigma_1)
    = p + 2\, v^2 \;, \quad {\textstyle{\frac{1}{2}}} \mathrm{Tr}
    (X \sigma_2 X \sigma_2) = p + 2\, w^2 \;.
\end{equation}
A notable feature here is that the variable $u$ occurs only in the
factor $\mathrm{e}^{-\epsilon N \mathrm{Tr}\, X} = \mathrm{e}^{- 2
\epsilon N \sqrt{p + v^2 + w^2} \cosh u}$. Thus the integral over $u$
for fixed $\beta := 2\,\epsilon N \sqrt{p + v^2 + w^2} \not= 0$ can
be carried out and yields the hyperbolic Bessel function
\begin{equation}\label{eq:bessel}
    \int_0^\infty \mathrm{e}^{-\beta\cosh u}du = K_0(\beta)\;.
\end{equation}

Let us now show how the functions $\tilde{h}_{N,\,k}(\tau)$ are
expressed in terms of the Hermite polynomials $H_n(x)$ defined by
$H_n(x)\,\mathrm{e}^{-x^2}= (-1)^n\, d^n / dx^n\,\mathrm{e}^{-x^2}$.
By shifting and setting the variable to zero after differentiation we
can rewrite this definition as
\begin{equation}
    H_n(x) = \left( \frac{d^n}{dy^n} \,\mathrm{e}^{-y^2 + 2xy}\right)
    \bigg\vert_{y=0} = \frac{\mathrm{i}^n n!}{2\pi\mathrm{i}}\oint_{
    \mathrm{U}(1)} dy\; y^{-n-1} \mathrm{e}^{-2\mathrm{i}x y+ y^2}\;,
\end{equation}
where we have used Cauchy's formula $(d^n f / dy^n)(0) = (2\pi
\mathrm{i})^{-1} n! \oint_{\mathrm{U}(1)} f(y)\, y^{-n-1} dy\,$. We
ultimately want to take the limit $N \to \infty$. To get a good view
of the large-$N$ asymptotics we introduce the scaled Hermite
polynomials
\begin{equation}\label{eq:auxiliary}
    h_n(\tau) = (-1)^n\,(2\,\mathrm{e}/n)^{-n/2}\, n!^{-1}
    H_n(\tau / \sqrt{2n}) = \frac{\mathrm{i}^{-n}}{2\pi\mathrm{i}}
    \oint_{\mathrm{U}(1)} dy\; y^{-n-1} \mathrm{e}^{-\mathrm{i} \tau
    y + (n/2) (y^2 - 1)} \;.
\end{equation}
By a saddle-point computation of the $\mathrm{U}(1)$ integral, these
polynomials have the large-$n$ behavior
\begin{equation}\label{eq:asympt}
    h_n(\tau) \simeq (n\pi)^{-1/2} \cos(\tau + n\pi/2) \;.
\end{equation}
Comparing the integrals (\ref{eq:auxiliary}) and (\ref{eq:tilde-h})
we read off the relation
\begin{equation}
    \tilde{h}_{N,\,k}(\tau) = (1-a^2)^{k+n/2}\,
    (\mathrm{e}N/n)^{\,n/2}\, h_n\left(\tau
    \sqrt{\frac{n}{N(1-a^2)}}\,\right)\;, \qquad n = N-2-k\;.
\end{equation}

A further simplification of Eq.\ (\ref{eq:partfunc}) is now achieved
by the 3-term recursion formula
\begin{equation}\label{eq:3-term}
    N \tilde{h}_{N,2}(\tau) + \tau \tilde{h}_{N,1}(\tau) +
    (N-2)(1-a^2) \tilde{h}_{N,0}(\tau) = 0 \;,
\end{equation}
which results from partially integrating $(N(1-a^2)\, y - d/dy)\,
\mathrm{e}^{\frac{N}{2} (1-a^2) y^2} = 0$ against $\mathrm{e}^{-
\mathrm{i} \tau y}\, y^{-N+2} dy\,$. Using the identity
(\ref{eq:3-term}) to eliminate the $\tilde{h}_{N,2}$ term from
(\ref{eq:partfunc}) we arrive at
\begin{eqnarray}
    && Z_{-1} = \frac{2\,N^{N+2}}{\pi (N-1)!} \int_0^\infty dp\;
    p^{N-2}\, \mathrm{e}^{-N(1+a^2)p} \int_\mathbb{R} dv
    \int_\mathbb{R} dw \; \mathrm{e}^{2 \mathrm{i}N (v\, \mathrm{Re}
    z + w\, \mathrm{Im} z) - 2N (v^2 + a^2 w^2)} K_0(2\,\epsilon N
    \sqrt{p + v^2 + w^2}) \nonumber \\ &\times& \left( (2\,\mathrm{i}
    v + 2\, a^2 w + z_f^\ast) \tilde{h}_{N,1}(N z_f^\ast) +
    \big(z_f^\ast (2\, \mathrm{i} v + 2\, a^2 w + z_f^\ast) +
    p\, (1-a^4) - (1-N^{-1})(1-a^2)\big) \tilde{h}_{N,0}(N z_f^\ast)
    \right) . \label{eq:partfunc2}
\end{eqnarray}
This expression for the partition function is exact for all matrix
dimensions $N \ge 2\,$. (It is, however, false for $N = 1$ because
the superbosonization formula fails in that case; see the discussion
in \cite{LSZ,BEKYZ}.)

\subsection{Calculation of the $\log \epsilon$ term}

We now extract from the integral representation (\ref{eq:partfunc2})
the term which is singular in the limit $\epsilon \to 0\,$. For
$\epsilon \to 0$ we may replace the hyperbolic Bessel function
(\ref{eq:bessel}) by its leading logarithm,
\begin{equation}\label{eq:expand}
    K_0(\beta) \simeq - \log(\beta/2) =
    - \log\epsilon - \log\big( N\sqrt{p + v^2 + w^2}\, \big) \;,
\end{equation}
where we keep only the singular term $\log\epsilon$ for now. The
integrals over the variables $v$ and $w$ then become Gaussian with
mean values $\langle v \rangle = \frac{ \mathrm{i}}{2} \,
\mathrm{Re}\, z$ and $\langle w \rangle = \frac{ \mathrm{i}}{2a^2}\,
\mathrm{Im}\,z$ and variances $\mathrm{var} (v) = (4N)^{-1}$ and
$\mathrm{var}(w) = (4N a^2)^{-1}$. The remaining integral over $p$
after scaling $p \to N^{-1} (1+a^2)^{-1} p$ yields the gamma function
$\Gamma(N-1) = \int_0^\infty p^{N-2}\, \mathrm{e}^{-p}\, dp$ and a
similar term with $N-1$ replaced by $N$. Altogether we obtain
\begin{equation}
    \lim_{\epsilon \to 0} |\log \epsilon\,|^{-1} Z_{-1} =
    \frac{N^2\,\mathrm{e}^{-\frac{N}{2}\mathrm{Re}^2(z) -
    \frac{N}{2}\mathrm{Im}^2(z/a)}}{(N-1)(1+a^2)^{N-1} a} \,
    (z_f^\ast - z^\ast) \big(\tilde{h}_{N,1} (N z_f^\ast) +
    z_f^\ast \,\tilde{h}_{N,0} (N z_f^\ast) \big) \;.
\end{equation}
In view of the asymptotic behavior (\ref{eq:asympt}) it is clear that
this will tend to a good limit for $N \to \infty$ when the product $N
z_f^\ast$ is kept fixed. Inserting the definition of the polynomials
$\tilde h_{n,\,k}$ and using the recursion relation
\begin{equation}\label{Hrecur}
    H_{n+1}(x) = 2 x H_n(x) - 2 n H_{n-1}(x)
\end{equation}
with $n = N-2\,$, we find the simplified expression
\begin{equation}\label{eq:partfunc1}
    \lim_{\epsilon \to 0} |\log \epsilon\,|^{-1} Z_{-1} =
    a^{-1} (z^*_f - z^*)\, \mathrm{e}^{-\frac{N}{2}
    \mathrm{Re}^2(z) -\frac{N}{2}\mathrm{Im}^2(z/a)}\,
    C_N(a)\, (-1)^{N} H_{N-1}(b z_f^*)\;,
\end{equation}
where the normalization constant $C_N(a)$ and scale factor $b \equiv
b(a)$ are given by
\begin{equation}\label{cnal}
    C_N(a) = (2b\,(1+a^2))^{-N+1} \frac{N^N}
    {(N-1)!} \;, \qquad b = \sqrt{\frac N{2(1-a^2)}} \;.
\end{equation}
This closed-form expression for the $\log \epsilon$ contribution is
still exact for all matrix dimensions $N \ge 2\,$.

Let us check that this result is consistent with the expression
(\ref{Zm1SUSY}) obtained in the large-$N$ limit. For that we observe
that the Hermite polynomials $H_n(x)$ for $\sqrt{n}\,x$ fixed and $n
\to \infty$ are asymptotic to
\begin{equation}\label{hnasym}
    (-1)^n H_n(x) \simeq \sqrt{2}\, (2\,n/\mathrm{e})^{n/2}
    \cos (\sqrt{2\,n}\,x + n\pi/2) \;.
\end{equation}
Recalling that in the microscopic limit we send $N\to \infty$ while
keeping $N a^2$, $N z^\ast$ and $N z_f^\ast$ fixed, we then see that
the microscopic limit given in (\ref{Zm1SUSY}) is precisely
reproduced.

\subsection{Contribution of order $\epsilon^0$}\label{sect4b}

We now set $z_f^\ast = z^\ast$ and compute the $\epsilon^0$
contribution to the partition function. This contribution is given by
the $\log (p+v^2+w^2)$ term in the expansion (\ref{eq:expand}) of the
hyperbolic Bessel function. (From the preceding section we know that
the constant terms in the expansion of $K_0(\beta)$ yield zero for
$z_f^\ast = z^\ast$.) To facilitate the computation, we write
\begin{equation}
    -2\log\sqrt{p+v^2+w^2}=\lim_{\delta\to 0}\left(\int_\delta^\infty
    \frac{dr}{r}\,\mathrm{e}^{-r(p+v^2+w^2)}+\log\delta+\gamma\right),
\end{equation}
where $\gamma$ is Euler's constant. The singular constant $\log
\delta\,$ and $\gamma$ can be dropped as they make no contribution
for $z_f^* = z^*$. The integrals over $v$, $w$, and $p$ can then be
carried out as before, and the resulting limit $\delta \to 0$ exists.
The order $\epsilon^0$ contribution is thus given by
\begin{eqnarray}\label{eq:partfunc4}
    Z_{-1}(z^*|z,z^*;a) &=& \big({\textstyle{\frac{N}{2}}}
    (1-a^2)\big)^{\frac{N}{2}-1} \int_0^\infty dr
    \;\frac{\mathrm{e}^{-(2N+r)^{-1} \mathrm{Re}^2(Nz)-(2Na^2+r)^{-1}
    \mathrm{Im}^2(Nz)}}{(1+a^2+N^{-1}r)^{N-1} \sqrt{(2N+r)(2Na^2+r)}}
    \nonumber\\ &\times& (-1)^N \frac{N^3}{N!}
    \left(\frac{N}{2b} \Big( \frac{\mathrm{Re}\,z}
    {2N+r} -\frac{\mathrm{i}\, \mathrm{Im}\,z}{2Na^2 + r} \Big) H_{N-1}
    (b z^*)-\frac{(N-1)(1-a^2)}{N(1+a^2)+r}\,H_{N-2}(b z^*)\right)\;.
\end{eqnarray}
This is the exact finite-$N$ result for the bosonic partition
function. Using the orthogonal polynomial approach it can also be
obtained from the Cauchy transform of the fermionic partition
function \cite{gerpr}.

Let us take once again the microscopic limit ($N\to\infty$, with
$Nz^\ast$, $Nz$ and $Na^2$ fixed). Doing so in Eq.\
(\ref{eq:partfunc4}) we get
\begin{equation}
    Z_{-1} = \frac{\mathrm{e}^{\frac{1}{2} N -
    \frac{3}{2} N a^2}}{\sqrt{2\pi}} \int_0^\infty dr \;
    \frac{\mathrm{e}^{-(2Na^2+r)^{-1} \mathrm{Im}^2(Nz)}}
    {\mathrm{e}^r \sqrt{2Na^2+r}}\left( \frac{\mathrm{i}\,
    \mathrm{Im}(Nz)}{2Na^2 + r} \sin(N z^\ast + N\pi/2)
    + \cos(Nz^\ast + N\pi/2) \right).
\end{equation}
To check this result we invoke the following identity:
\begin{equation}\label{eq:identity}
    \sqrt{\pi} \int_0^\infty dr\; \frac{\mathrm{e}^{-N^2\,
    \mathrm{Im}^2(z)/(2Na^2+r)}}{\mathrm{e}^r \sqrt{2Na^2+r}}
    = \int_\mathbb{R} \frac{dq}{1+q^2}\; \mathrm{e}^{-2N a^2
    q^2 + 2\mathrm{i}N q\,\mathrm{Im}\,z}\;,
\end{equation}
and a second identity of the same kind which is obtained by
differentiating both sides of (\ref{eq:identity}) with respect to
$\mathrm{Im}\,z\,$. Using Euler's formula $\cos\theta + \mathrm{i}
\sin\theta = \mathrm{e}^{\mathrm{i} \theta}$ we then immediately
recover Eq.\ (\ref{super}).

\section{Calculation of $Z_{-1}$ using complex orthogonal polynomials}
\label{sec:OP}

In this section we first derive the partition function with one
bosonic quark from the Cauchy transform of orthogonal polynomials. In
this approach no regularization procedure is required. For comparison
with the $\sigma$-model we also compute the partition function with
one fermionic quark and two conjugate bosonic quarks, which diverges
as $\log \epsilon\,$.

To apply the method of complex orthogonal polynomials to the FKS
model we first express the Gaussian probability distribution for $H$
and $A$ given in Eq.\ (\ref{PHA}) in terms of the eigenvalues $z_k$
of $H+A$ and $z_k^*$ of $H-A\,$. The joint distribution for the
eigenvalues was calculated in \cite{sommers-rev}. Including the
expression for the exponent, which follows from the decomposition
\begin{equation}
    {\rm Tr}HH^\dagger + \frac 1{a^2}{\rm Tr}AA^\dagger =
    \frac {1+1/a^2}2 {\rm Tr}\,(H+A)(H^\dagger+A^\dagger) +
    \frac{1-1/a^2}4{\rm Tr} \big( (H+A)^2 + (H^\dagger +
    A^\dagger)^2 \big) \;,
\end{equation}
the joint eigenvalue distribution function is given by
\cite{sommers-rev}
\begin{equation}
    P(\{z_k,z_k^*\}) = C \prod_{k<l} |z_k - z_l|^2 \mathrm{e}^{-
    \frac{N}{4}(1+1/a^2) \sum_k |z_k|^2 -\frac{N}{8} (1-1/a^2)
    \sum_k( z_k^2 +z_k^{*\,2})}.
\end{equation}
The partition functions (and correlation functions) of the FKS model
can now be derived by means of the method of complex orthogonal
polynomials with polynomials $p_n(z)$ defined through
\begin{equation}
    \int d^2z \; w(z,z^*;a)\, p_k(z)\, p_l(z^*) = r_k \delta_{kl}
\end{equation}
and weight function given by
\footnote{In terms of the parameter $\tau = (1-a^2)/(1+a^2)$
introduced in \cite{sommers} the weight reads $w(z,z^*;\tau) =
\exp[-N|z|^2/(2(1-\tau))+ N\tau/(1-\tau)(z^2 +z^{*\,2})].$}
\begin{equation}
    w(z,z^*;a) = \mathrm{e}^{-\frac N4(1+1/a^2)|z|^2-
    \frac{N}{8} (1-1/a^2)(z^2 +z^{*\,2})}.
\end{equation}
These polynomials, which have been known for some time
\cite{itzykson,forrester,FKS}, are given by
\begin{equation}
    p_n(z) \sim H_n(b z), \qquad b = \sqrt{\frac N{2(1-a^2)}}\;,
\end{equation}
where $H_n$ are the Hermite polynomials. The $p_k$ are in monic
normalization with respect to $z\,$. The leading $a$-de\-pen\-dence
of $r_k$ for large $k$ is given by
\begin{equation}
    r_k \sim \mathrm{e}^{\frac{1}{2} k a^2}.
\end{equation}
In this section, we will not keep track of numerical and
$a$-dependent prefactors.

General expressions for partition functions in terms of complex
orthogonal polynomials have been given in \cite{B1,B2,AP}. Below we
derive the explicit expressions for the microscopic limit of
$Z_{-1}(z^*|z,z^*;a)$ and $Z_{-1}(z_f^*|z,z^*;a)$.

\subsection{The partition function $Z_{-1}(z^*|z,z^*;a)$}

The partition function with one boson, $Z_{-1}(z^*|z,z^*;a)$, can be
expressed as a Cauchy transform \cite{B1,B2,AP}
\begin{equation}
    Z_{-1}(z^*|z,z^*;a) = \frac{1}{r_{N-1}} \int d^2z' \;
    w(z',{z'}^*;a)\, p_{N-1}({z'}^*) \frac{1}{z-z'} \;.
\end{equation}
In the microscopic limit where $Nz$ and $Na^2$ are kept fixed for
$N\to \infty$, the Hermite polynomials can be replaced by their
asymptotic limit (\ref{hnasym}) and the weight function reduces to
\begin{equation}
    w(z,z^*;a) \sim \mathrm{e}^{-\frac{N\mathrm{Im}^2(z)}{2a^2}}\;.
\end{equation}
We thus find
\begin{equation}
    Z_{-1}(z^*|z,z^*;a) \sim (-\mathrm{i})^{N-1} \mathrm{e}^{-
    \frac{3}{2} N a^2} \int dx'dy'\; \mathrm{e}^{-\frac{N{y'}^2}
    {2a^2}} \frac{\mathrm{e}^{\mathrm{i}N{z'}^*} + (-1)^{N-1}
    \mathrm{e}^{-\mathrm{i}N{z'}^*}} {z-x'-\mathrm{i}y'}\;.
\end{equation}
The integral over $x'$ can be performed by a contour integration,
whereas the remaining integral over $y'$ can be expressed in terms of
the complementary error function. This leads to the expression
\begin{eqnarray}
    Z_{-1}(z^*|z,z^*;a) &\sim& (-\mathrm{i})^{N} \mathrm{e}^{-
    \frac{3}{2} Na^2}\int dy'\; \mathrm{e}^{-\frac{N{y'}^2}{2a^2}}
    \left( \mathrm{e}^{\mathrm{i}Nz} \mathrm{e}^{2Ny'}\theta(-y'+
    \mathrm{Im} z) + (-1)^N \mathrm{e}^{-\mathrm{i}Nz}\mathrm{e}^{-
    2Ny'} \theta(y'-\mathrm{Im} z)\right) \nonumber \\ &\sim&
    \mathrm{e}^{\frac{1}{2} Na^2} \mathrm{e}^{\mathrm{i}N(z+
    \pi/2)}\,\mathrm{erfc}\left(\frac{2Na^2- N\mathrm{Im} z}
    {\sqrt{2N}a}\right) + \mathrm{e}^{\frac{1}{2} Na^2}
    \mathrm{e}^{-\mathrm{i}N(z+\pi/2)}\,\mathrm{erfc}\left(
    \frac{2Na^2 + N\mathrm{Im} z}{\sqrt{2N}a}\right) \;,
\end{eqnarray}
in agreement with (\ref{Z1bosSUSY}).

\subsection{The partition function $Z_{-1}(z_f^*|z,z^*;a)$}

In \cite{AOSV} it was shown that the singular part of the chiral
random matrix partition function with a pair of conjugate bosonic
quarks and $N_f$ fermionic flavors factorizes. The same reasoning can
be applied to the FKS model resulting in
\begin{equation}
    Z^{(N)}_{-1}(z_f^*|z,z^*;a)\sim (z_f^*-z^*)
    \,Z^{(N-1)}_1(z_f^*;a)\, Z^{(N)}_{\rm pq-bos}(z,z^*;a)
    + \mathcal{O}(\epsilon^0), \label{AOSV-fact}
\end{equation}
where the phase quenched bosonic partition function, $Z^{(N)}_{\rm
pq-bos}\,$, is defined in (\ref{pq-bos}) and $Z^{(N-1)}_1$ is the
partition function with one fermionic flavor. $Z^{(N)}_{\rm pq-bos}$
is given by the weight function times $\log \epsilon$
\cite{SplitVerb2,AOSV}
\begin{equation}
    Z_{\rm pq-bos}(z,z^*;a) = |\log\epsilon| \
    \frac{w(z,z^*;a)}{r_{N-1}} \;, \label{AOSV-43}
\end{equation}
and the $N_f=1$ theory can be expressed in terms of the orthogonal
polynomials \cite{GernotQCD3}
\begin{equation}
    Z^{(N-1)}_{1}(z_f^*;a) \sim p_{N-1}(z_f^*)\;.
\end{equation}
Inserting this result and (\ref{AOSV-43}) into (\ref{AOSV-fact}) we
obtain
\begin{equation}
    Z^{(N)}_{-1}(z_f^*|z,z^*;a) \sim (z_f^* -z^*) w(z,z^*;a)
    \,\mathrm{e}^{-\frac{3}{2} Na^2} H_{N-1}(b z_f^*)\,
    \log \epsilon + \mathcal{O}(\epsilon^0) \;,
\end{equation}
in agreement with the exact finite-$N$ result in
(\ref{eq:partfunc1}). The factor $\exp(-3Na^2/2)$ results from a
factor $\exp(-Na^2)$ from $r_{N-1}$ and a factor $\exp(-Na^2/2)$ from
the ratio of $p_{N-1}$ (in monic normalization) and $H_{N-1}\,$. In
the microscopic limit this results in
\begin{equation}
    Z_{-1}(z_f^*|z,z^*;a) \sim (z_f^* -z^*)\,\mathrm{e}^{-\frac{3}{2}
    Na^2} \mathrm{e}^{-N\,\mathrm{Im}^2(z)/(2a^2)}\sin(Nz_f^* + N\pi/2)
    \,|\log \epsilon| + \mathcal{O}(\epsilon^0),
\end{equation}
in agreement with the result (\ref{Zm1SUSY}) obtained earlier in this
paper.

\section{Conclusions}

We have analyzed the bosonic partition function of a Hermitian random
matrix model deformed by a nonhermitian random matrix model. We have
shown that the microscopic limit of the partition function can be
obtained by essentially using symmetry arguments only. There are,
however, several subtleties that deserve attention. First, the
partition function has to be regularized by multiplication with a
conjugate bosonic and conjugate fermionic determinant. Second,
because fermionic degrees of freedom are present, two inequivalent
saddle-point manifolds have to be taken into account. Third,
convergence of the partition function leads to the boson-boson block
of the manifold of the Goldstone degrees of freedom being a
noncompact subset of the set of positive definite matrices.

The main advantage of the symmetry approach is that it gives a clear
view at universality. Goldstone modes, which are separated from the
rest of the excitation spectrum by a mass gap, decouple in the
microscopic limit, and their mutual interactions are completely
determined by the symmetries and the pattern of symmetry breaking of
the microscopic partition function. This means that our results for
the FKS model are valid for the microscopic limit of any model with
the same symmetries and a mass gap.

To obtain results for finite-size matrices one has to perform a
detailed calculation. We have presented results using two different
methods: the superbosonization method and the complex orthogonal
polynomial method. The disadvantage of the orthogonal polynomial
method is that universality is not manifest at all stages of the
calculation. In the superbosonization method the universal partition
function is obtained after integrating out the massive modes, which
is a trivial step when the proper coordinates are used. The
orthogonal polynomial approach, which is applicable to invariant
random matrix models, has as its main advantage that it can be
generalized in a straightforward way to any number of flavors. We
have also performed the calculation using a hybrid method where the
four-fermion term is decoupled by means of the Hubbard-Stratonovich
transformation. Since this calculation did not provide additional
insights, we have refrained from presenting it in this paper.

For the present problem, the superbosonization approach does not have
a clear advantage over the hybrid method, but in general we expect
that it will be simpler to integrate out the massive modes if
fermions and bosons are treated in a unified way. We also wish to
stress that a major advantage of the superbosonization method is that
it can deal with nongaussian probability distributions. Such
distributions have important applications in, e.g., quantum gravity
and growth phenomena. However, nongaussian perturbations do not
affect the universal results obtained in the microscopic limit. These
are determined by symmetries and can be derived from symmetry
arguments alone as we have shown in this paper. \vspace{2mm}

\noindent{\sl Acknowledgments.} We wish to thank Gernot Akemann and
Poul Damgaard for stimulating discussions. This work was supported by
U.S. DOE Grant No. DE-FG-88ER40388 (JV), the Carlsberg Foundation
(KS), the Villum Kann Rassmussen Foundation (JV), the Danish National
Bank (JV), and the Deutsche Forschungsgemeinschaft, SFB/TR 12 (MRZ).


\end{document}